\documentclass[aps,prb,twocolumn,letterpaper,superscriptaddress,showpacs]{revtex4-1}
\usepackage{keyval}%
\usepackage{graphicx}
\usepackage{dcolumn}
\usepackage{bm}
\usepackage{color}
\usepackage{amsmath,amssymb}

\begin{document}

\title{Partitioning interatomic force constants for first-principles phonon calculations: Applications to NaCl, 
PbTiO$_3$, monolayer CrI$_3$, and twisted bilayer graphene}
\author{Chi-Cheng Lee}
\affiliation{Department of Physics, Tamkang University, Tamsui, New Taipei 25137, Taiwan}%
\affiliation{Research Center for X-ray Science, College of Science, Tamkang University, Tamsui, New Taipei 25137, Taiwan}%
\author{Chin-En Hsu}
\affiliation{Department of Physics, Tamkang University, Tamsui, New Taipei 25137, Taiwan}%
\author{Hung-Chung Hsueh}
\affiliation{Department of Physics, Tamkang University, Tamsui, New Taipei 25137, Taiwan}%
\affiliation{Research Center for X-ray Science, College of Science, Tamkang University, Tamsui, New Taipei 25137, Taiwan}%
\date{\today}

\begin{abstract}
First-principles phonon calculations have been widely performed for studying vibrational properties of condensed matter, 
where the dynamical matrix is commonly constructed via supercell force-constant calculations or the linear response approach.
With different manners, a supercell can be introduced in both methods. Unless the supercell is large enough, the
interpolated phonon property highly depends on the shape and size of the supercell and the imposed periodicity could give unphysical results 
that can be easily overlooked. Along this line, the concept of partition of force constants is discussed, and addressed by 
NaCl, PbTiO$_3$, monolayer CrI$_3$, and twisted bilayer graphene as examples for illustrating the effects of the imposed 
supercell periodicity. To diminish the unphysical effects, a simple method of partitioning force constants, which relies only on the translational 
symmetry and interatomic distances, is demonstrated to be able to deliver reasonable results. 
The partition method is also compatible with the mixed-space approach for describing LO-TO splitting. 
The proper partition is especially important for studying moderate-size systems with low symmetry, such as two-dimensional materials on substrates, 
and useful for the implementation of phonon calculations in first-principles packages using atomic basis functions,
where symmetry operations are usually not applied owing to the suitability for large-scale calculations.
\end{abstract}
  
\maketitle

\section{Introduction}

First-principles phonon calculations based on density functional theory\cite{Hohenberg,Kohn} have been successful in describing various vibrational 
and thermodynamic properties of condensed matter, owing to the accurate description of total energy surfaces against atomic 
displacements.\cite{RevModPhys.73.515,Wang2016}
Thanks to the computational power available nowadays, not only is the amount of phonon calculations from first principles rapidly increasing 
but also building phonon databases via high-throughput calculations becomes essential for large-scale exploration of new materials.\cite{Petretto} 
The underlying theoretical foundation for the most performed phonon calculations relies on the harmonic approximation, which makes 
the physics easily accessible, and can be straightforwardly extended for exploring anharmonic effects,\cite{PhysRevLett.40.950,Cowley_1968} 
including the progress in the field of information science, for example, the compressive sensing method,\cite{PhysRevLett.113.185501} 
in studying phonons with sufficient training data.
In the implementation, the dynamical matrices, whose eigenvalues deliver the phonon dispersion, can be constructed via the real-space force constants 
obtained directly from the supercell calculations by displacing the atoms from equilibrium 
through the finite-difference method\cite{Kresse_1995,Ackland_1997,PhysRevLett.78.4063,Parlinski} 
or the self-consistent calculations based on the density functional perturbation theory (DFPT) 
in the framework of linear response theory,\cite{PhysRevB.55.10337,PhysRevB.55.10355} 
where the response, and therefore the dynamical matrix, can be calculated at any arbitrary wave vector ($\vec{q}$) accurately.

As long as the harmonic approximation is valid, both the supercell force-constant and linear response approaches should in principle reach the same result
on the condition that the supercell is large enough allowing the forces decaying into negligible values away from the displaced atom 
and the dynamical matrices are calculated by DFPT at all the studied $q$ points, respectively.
However, a large-scale supercell calculation, especially for systems having imperfection, such as defects 
and the presence of surfaces, could be difficult. For the case the primitive-cell size is large, the difficulty still remains in the adoption of 
linear response approach since the required self-consistent calculations are time-consuming. In addition, the implementation with 
complicated functionals in the framework of DFPT could be challenging.\cite{Arrigo}
To reduce the computational effort for the DFPT calculation where getting a $q$ solution is still affordable, 
the dynamical matrices can be calculated only at the $q$ grids needed for the Fourier transform to obtain the real-space force constants. 
Note that the dynamical matrix can be considered as the reciprocal-space force constants regardless of the atomic mass. In this context, 
a denser $q$-mesh corresponds to a larger supercell and can have more complete long-range interatomic force constants that can be used for 
interpolating full phonon dispersion with higher accuracy.  

Although the linear response approach can deliver accurate results without resorting to the Fourier transformation, 
which can be considered as the exact theoretical solutions for the harmonic approximation, 
the supercell force-constant calculations have still been extensively adopted for studying phonon properties. This is because the concept of force constants 
in real space is intuitive. Meanwhile, the implementation of DFPT\cite{SHANG201726} 
is not available in most of the first-principles packages using atomic basis functions,\cite{Wang2016}
which are efficient for large-scale calculations.\cite{0953-8984-14-11-302,GILLAN200714,BLUM20092175,hutter2014cp2k,quantumwise,Ozaki}.
Moreover, the constructed dynamical matrices via the supercell force-constant approach are expected to be consistent with 
those obtained by DFPT for the wave vectors commensurate with the adopted supercell. 
In this case, the requirement for having negligible forces away from the displaced atom becomes irrelevant. Although the commensurability offers 
a luring way to investigate phonons at the high-symmetry points in the Brillouin zone, the number of those $q$ vectors is quite limited.
To describe thermodynamic properties via phonon density of states, the phonon dispersion in the entire Brillouin zone is needed. 
Recently, there are several important efforts that have led to great improvements of determining phonon properties associated with the incommensurate $q$ vectors, 
such as the inclusion of a denser $q$-mesh only in the region where more accurate Fourier interpolations are required,\cite{PhysRevB.77.024309} 
the compressive sensing lattice dynamics (CSLD) method\cite{PhysRevB.100.184309} for better fitting phonon dispersion by modeling the force constants,
and the mixed-space method\cite{mixed,PhysRevB.85.224303} for better interpolating the frequency splitting between the longitudinal optical 
and transverse optical (LO-TO) modes due to the dipole-dipole interactions. 
Very recently, higher-order multipolar interactions between atoms are found to be important for improving the acoustic phonon dispersion around 
the Brillouin zone center in piezoelectric materials.\cite{Stengel}
Along this line, we will address an issue of partitioning force constants, 
which can improve the interpolated phonon properties associated with the incommensurate $q$ vectors while keeping those with the commensurate ones unaffected,
for the phonon calculations using the supercell force-constant approach.

The paper is organized as follows. The dynamical matrix, including the non-analytic contribution, and how the mixed-space method can be used to
interpolate the phonon dispersion with LO-TO splitting using the supercell force-constant method are given in Sec.~\ref{sec:dynamicalmatrix}. 
The concept of partition of interatomic force constants and 
the related formulae are discussed in Sec.~\ref{sec:partition} A. In Sec.~\ref{sec:partition} B, a useful partition method is proposed.
The applications of the proposed partition method to NaCl, PbTiO$_3$, monolayer CrI$_3$, and twisted bilayer graphene are presented in 
Sec.~\ref{sec:application}. Finally, a summary is given in Sec.~\ref{sec:summary}.

\section{Dynamical matrix}
\label{sec:dynamicalmatrix}

The dynamical matrix, $D$, whose eigenvalues give the phonon frequencies (more precisely, the squares), can be constructed via 
the interatomic force constant, $C$: 
\begin{equation}
D^{st}_{\alpha\beta}(\vec{q})=\frac{1}{\sqrt{M_sM_t}}\sum_{l=1}^{N}C^{st}_{\alpha\beta}(0,\vec{R}_l)e^{i\vec{q}\cdot\vec{R}_l},
\label{eq:eq01}
\end{equation}
where $M_s$ and $M_t$ denote the mass of atoms $s$ and $t$ belonging to the primitive unit cell, respectively, 
and $\alpha$ and $\beta$ index the Cartesian components ($x$, $y$, and $z$).
$\vec{R}_l$ labels the primitive lattice vector and $l$ sums over the primitive unit cells in the supercell, so $N$ indicates
the number of the primitive unit cells in the supercell.

The non-analytic contribution, $^{na}D$, which should be added to Eq.~\ref{eq:eq01} to account for the LO-TO splitting 
in the long-wavelength limit ($\vec{q}\rightarrow 0$),\cite{Cochran} is expressed as 
\begin{equation}
^{na}D^{st}_{\alpha\beta}=\frac{4\pi e^2}{V}\frac{(\bm{q}\cdot\bm{Z^*}_s)_\alpha(\bm{q}\cdot\bm{Z^*}_t)_\beta}{\bm{q}\cdot\bm{\epsilon}^{\infty}\cdot\bm{q}},
\label{eq:eq02}
\end{equation}
where $V$ denotes the volume of the primitive unit cell. $Z^*$ and $\epsilon^{\infty}$ denote the Born effective charge 
and the high-frequency dielectric constant,
respectively. Since the expected weight of the non-analytic contribution is 1 at $\Gamma$ (more precisely, at $\vec{q}\rightarrow 0$) 
and 0 at the commensurate $q$ points, an adjustable damping function, for example, $e^{-aq^2}$, can be introduced to the weight for interpolating
the phonon dispersion.\cite{PhysRevLett.81.3298} However, it should be noted that a Gaussian function cannot be entirely reduced to zero by a finite $q$. 
On the other hand, the weight has been chosen as the factor, 
\begin{equation}
f(\vec{q})=\frac{1}{N}\sum_{l}^{N}e^{i\vec{q}\cdot\vec{R}_l},
\label{eq:eq03}
\end{equation}
in the mixed-space method,\cite{mixed,PhysRevB.85.224303} which guarantees the required values of 1 and 0 at $\Gamma$ and the commensurate $q$ points, respectively.
As a result, the term,
\begin{equation}
\frac{4\pi e^2}{NV}\frac{(\bm{q}\cdot\bm{Z^*}_s)_\alpha(\bm{q}\cdot\bm{Z^*}_t)_\beta}{\bm{q}\cdot\bm{\epsilon}^{\infty}\cdot\bm{q}},
\label{eq:eq04}
\end{equation}
can be added together with the force constant to share the same phase factor, $e^{i\vec{q}\cdot\vec{R}_l}$, described in Eq.~\ref{eq:eq01}. 
The mixed-space method has been widely implemented in most of the phonon packages adopting the supercell force-constant 
approach.\cite{wang2014yphon,li2014shengbte,chernatynskiy2015phonon,togo2008first,tadano2014anharmonic}

To construct the dynamical matrix of a system, each primitive-cell atom in the supercell needs to be displaced, 
and the complete force constants can be obtained from the $3\times N_{atom}$ supercell calculations, where $N_{atom}$ denotes the number of atoms in the 
primitive unit cell. For a system with low symmetry, the required $3\times N_{atom}$ calculations cannot be reduced significantly and the 
so-called supercell could be the primitive unit cell itself, for example, in the case of studying the slab system with a surface reconstruction  
where the primitive unit cell is large enough for exploring desired phonon properties and doubling the in-plane unit cell is already unfavorable
for computation due to the thickness of the slab. One of the good strategies for studying this system is to fit the forces using 
the CSLD method\cite{PhysRevB.100.184309} by performing a smaller number ($< 3\times N_{atom}$) of supercell calculations.
In this study, we focus on how to improve the obtained force constants from a supercell that cannot be enlarged significantly due to the  
computational cost but is not large enough allowing the forces decaying into negligible values away from the displaced atom.

\section{Partition of force constants}
\label{sec:partition}

In the supercell force-constant calculations, the presence of the periodic images of atoms results in superposed forces in the supercell.
For a system where the decaying behavior of the interatomic force constants is known, the superposed force constants can be
partitioned into the individual contributions of the images. However, to analyze the decaying behavior of the force constants 
is challenging and might require larger-scale supercell calculations prior to the phonon calculations. 
As the supercell size increases, the effect of the superposed forces on the interpolated phonon dispersion
becomes less significant. However, without having a supercell that is immune to giving unphysical results due to the shape and size, 
a proper partition of force constants is suggested. In Sec.~\ref{subsec:concept}, the concept of partitioning force constants and a traditional 
partition method are discussed. Subsequently, a useful partition method is introduced in Sec.~\ref{subsec:bydistance}.

\subsection{Concept and formulation}
\label{subsec:concept}

\begin{figure}[tbp]
\includegraphics[width=1.00\columnwidth,clip=true,angle=0]{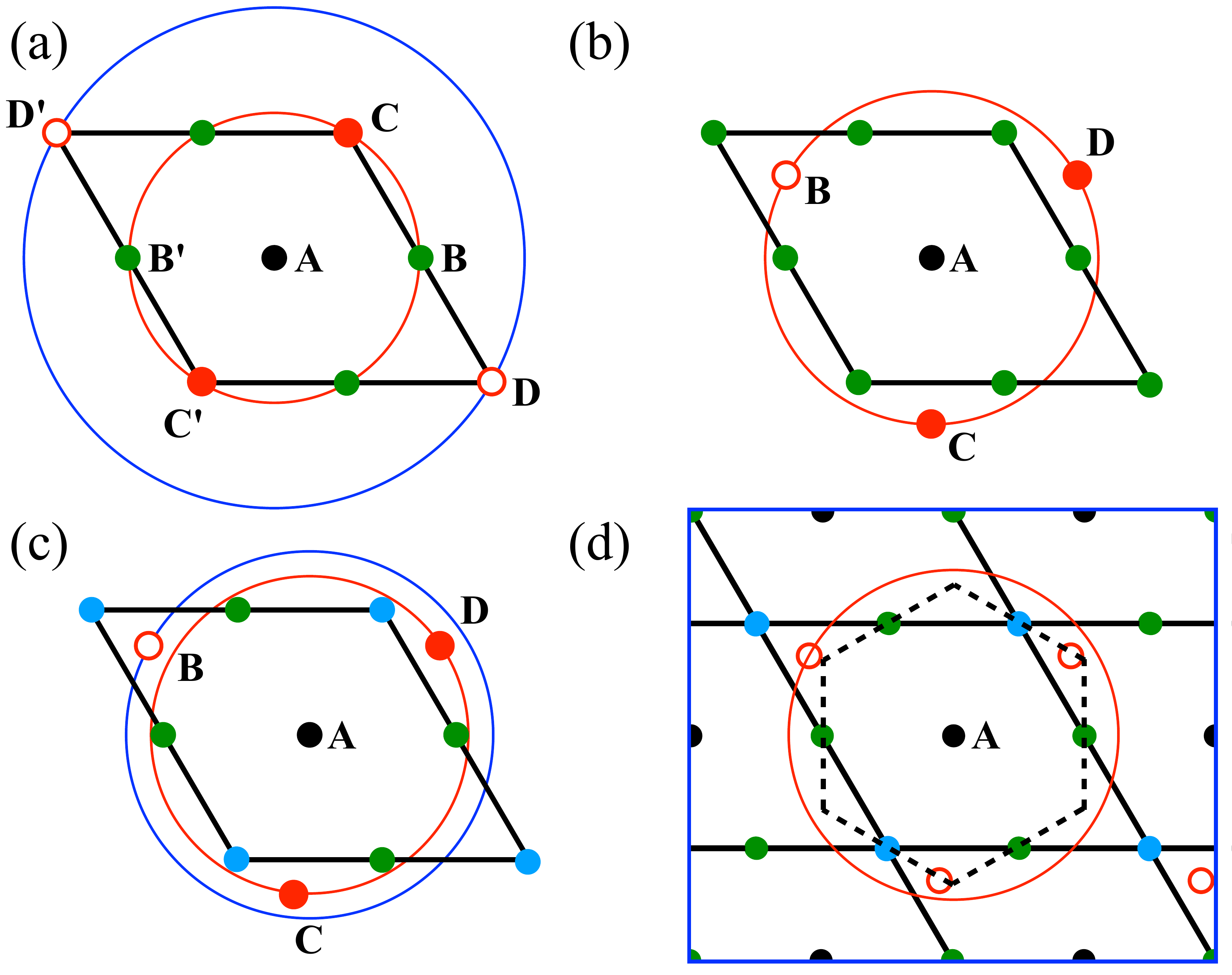}
\caption{(a) A schematic two-dimensional structure with the reference atom A plotted at the supercell center. (b) A hollow-site adatom B is introduced and
its periodic counterparts C and D are also presented. (c) Similar to (b), adatom B is introduced but deviates from the perfect hollow-site position,   
and its periodic images, C and D, are closer to atom A. (d) Another choice of supercell (blue rectangle) of the structure shown in (c) is presented together 
with the Wigner–Seitz cell (dashed lines). All of the guiding circles are centered at A.
}
\label{fig:model}
\end{figure}

A model structure of two-dimensional material is presented in Fig.~\ref{fig:model}(a) to illustrate the concept of partition of force constants. 
For clarity, the substrate is not shown and the reference atom A is plotted at the center of the unit cell, which is always possible owing to 
the periodic boundary condition. 
It can be observed that there are four periodic repeats of the corner atom  
and two pairs of periodic repeats at the boundary. 
Obviously, there is only one force that can be obtained for each set of periodic atoms after displacing atom A. 
In the literature, it was suggested that the force can be divided by the number of the periodic atoms 
locating at the supercell boundary.\cite{Ackland_1997,Parlinski} This traditional partition method,
dividing the force with an equal amount into the periodic atoms does provide a more symmetric distribution of the force constants and therefore 
a better interpolation of phonon dispersion for the non-commensurate $q$ vectors. 
For example, the biased A-B and A-C force constants 
are divided by two for getting the A-B and A-B$^\prime$ force constants and four for A-C, A-C$^\prime$, A-D, and A-D$^\prime$ force constants, 
respectively. Note that the partition is also needed for the interpolation procedure in the linear response approach, where the real-space force
constants are obtained via the Fourier transform. 

It is interesting to reformulate the dynamical matrix, $D^{st}_{\alpha\beta}(\vec{q})$, by the partitioned force constant $C^\prime$
together with the associated weight $W$: 
\begin{align}
&D^{st}_{\alpha\beta}(\vec{q})= \nonumber \\
&\frac{1}{\sqrt{M_sM_t}}\sum_{l=1}^{N}\sum_{L}^\infty 
C^{^\prime st}_{\alpha\beta}(0,\vec{R}_l+\vec{R}_L)e^{i\vec{q}\cdot(\vec{R}_l+\vec{R}_L)},  
\label{eq:eq05}
\end{align}
and   
\begin{equation}
C^{^\prime st}_{\alpha\beta}(0,\vec{R}_l+\vec{R}_L)=W^{st}_{\alpha\beta}(\vec{R}_l+\vec{R}_L)C^{st}_{\alpha\beta}(0,\vec{R}_l)
\label{eq:eq06}
\end{equation}
with 
\begin{equation}
\sum_L^\infty W^{st}_{\alpha\beta}(\vec{R}_l+\vec{R}_L)=1,
\label{eq:eq07}
\end{equation}
where $\vec{R}_L$ denotes the supercell lattice vector. In the traditional partition method, the weight is 0 for the atom $t$
outside the supercell and 1 inside the supercell. For the periodic atoms at the supercell boundary, the weight is 
obtained by dividing 1 with the number of the periodic atoms.
The beauty of the partition method is that 
the original dynamical matrices are unaffected at the commensurate vectors ($\vec{q}_{exact}$) because of $e^{i\vec{q}_{exact}\cdot\vec{R}_L}=1$. 

It should be addressed that there are abundant ways to satisfy Eq.~\ref{eq:eq07}. 
Regarding that the strength of A-D force constants should be 
smaller than the A-C ones due to the longer distance, another choice of partition is to divide the A-C 
force by two instead of four. At this step, the sum rule is still satisfied because the weight becomes 0.5, 0.5, 0, and 0 for the 
four atoms by treating the A-D and A-D$^\prime$ weight as 0 and the total sum is still 1. However, after applying rotational symmetry 
to further redistribute the force constants between atoms B and C, there is no guarantee that the phonon properties remain the same at the 
commensurate $q$ vectors since atoms B and C do not belong to the same periodic set. Although the symmetry operation does recover the
rotational symmetry if the system has the symmetry, it should be applied with caution;
otherwise, all the phonon properties could become the interpolated ones.
For the case of hexagonal symmetry, it has been pointed out that a special treatment is needed other than the traditional partition method.\cite{PhysRevB.55.10355}
In Sec.~\ref{sec:partition} B, we will focus on how to improve the interpolation while avoiding affecting the solutions at the      
commensurate $q$ vectors.

We now consider an adatom B, located at the hollow site as shown in Fig.~\ref{fig:model} (b). 
Clearly, the atoms C and D, which are the periodic counterparts of adatom B, are outside the supercell but cannot be neglected 
in analyzing the force constants associated with the reference atom A. For this case, 
atoms C and D can be found by applying the translational symmetry or rotational symmetry. 
For the supercell shown in Fig.~\ref{fig:model}(c), the adatom B is assumed to deviate from 
the perfect hollow-site position. 
If one considers only the force constants of the atoms inside the supercell, the force constants of the shorter-distance 
A-C and A-D counterparts will be overlooked. The resultant force constants can be unphysical depending on the strength of 
the A-B force constants, and the obtained vibrational modes could be incorrect even though the frequencies might still be close to the 
exact solutions. In this case, to obtain the periodic counterparts, what we need to do is to apply the translational 
symmetry, not the rotational symmetry. 

Alternatively, the supercell shown in Fig.~\ref{fig:model}(c) can be reshaped to the rectangular one shown in Fig.~\ref{fig:model}(d). 
However, the seemingly more symmetric distribution of the force constants cannot deliver the exact solutions
at the originally commensurate $q$ points, for example, the high-symmetry points in the Brillouin zone of the primitive unit cell,
since selecting a supercell is equivalent to the selection of the exact $q$ points. Moreover, the Wigner–Seitz cell 
shown in Fig.~\ref{fig:model}(d), which is a good choice for describing the force constants, explicitly rules out the 
contributions of adatom B and longer-distance corner atoms in the original supercell that could still play an important role.
Even for a cubic or tetragonal unit cell, as shown in Fig.~\ref{fig:model2}(b), 
the contribution of longer-distance corner atoms are also ruled out by the choice of Wigner–Seitz cell
in the case a reference atom slightly deviates from the center in the equilibrium structure. 
Therefore, it is timely to introduce a more flexible partition method. 

\subsection{Partition method by distance}
\label{subsec:bydistance}

\begin{figure}[tbp]
\includegraphics[width=1.00\columnwidth,clip=true,angle=0]{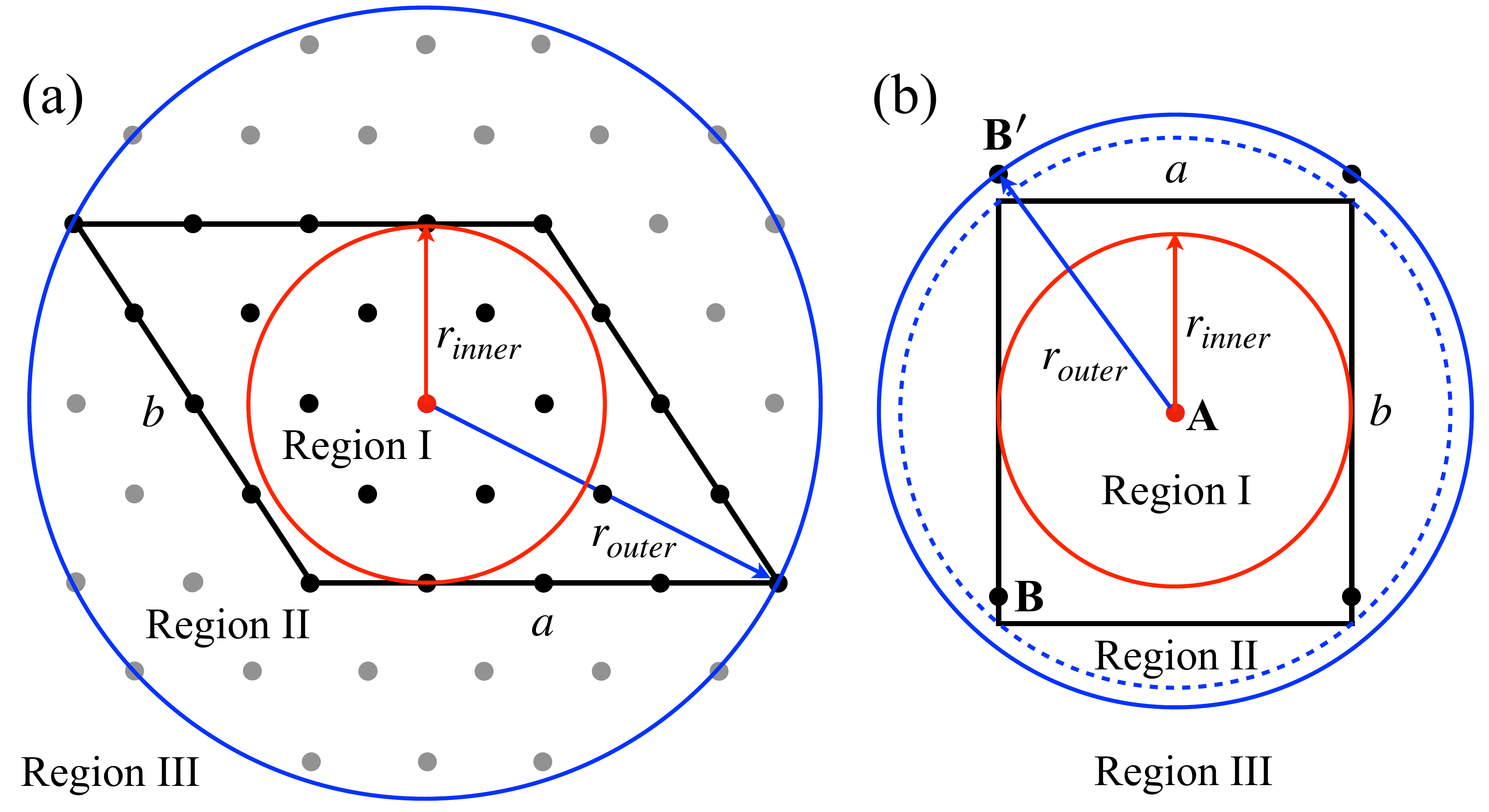}
\caption{ (a) A supercell contains 16 atoms. The reference atom has been shifted to the supercell center. 
The inscribed and circumscribed circles centered at the supercell center with their radii, $r_{inner}$ and $r_{outer}$, are presented, respectively. 
Three regions, I: $r<r_{inner}$, II: $r_{inner}\leq r\leq r_{outer}$, and III: $r>r_{outer}$, can be defined.
(b) Only two of the atoms, A and B, are shown in a supercell where the A-B interaction is considered to be important. 
A longer $r_{outer}$ is chosen to take into account the A-B$^\prime$ force constants. 
Note that the lattice constant $a$ might not be the same as $b$ in both (a) and (b), and the circles become spheres in three dimensions.
}
\label{fig:model2}
\end{figure}

We have introduced the weight for partitioning force constants in Eqs.~\ref{eq:eq06} and \ref{eq:eq07}.
Here, we propose a simple way to choose the weight based on the interatomic distance and translational symmetry. 
For a given supercell, for example, the one shown in Fig.~\ref{fig:model2} (a) or (b),
three regions, I: $r<r_{inner}$, II: $r_{inner}\leq r\leq r_{outer}$, and III: $r>r_{outer}$, can be defined based on the 
following consideration. The shorter-range interatomic force constants are expected to be more easily captured 
by the supercell force-constant method so it is desirable not to partition these force constants. 
The inner radius of a reference atom, $r_{inner}$, can then be chosen such that the force constants 
between the reference atom and the atoms in the region I ($r<r_{inner}$) remain intact. In this case, the corresponding weight 
is set to 1. For the short-range ingredients of force constants, for example, coming from a covalent bond between atoms responsible for higher frequencies,
partitioning such force constants could lead to incorrect results. For the long-range ingredients, for example,
coming from the dipole-dipole interactions, the superposed force constants in the region I provide an effective description for
the set of periodic images. Up to this point, the adopted force constants are consistent with those described by the
sophisticated supercell force-constant methods.

The rest of the interatomic force constants, especially those around the supercell boundary, are relevant to possibly giving unphysical
results. Considering the long-range force constants are small, 
a longer radius $r_{outer}$ can be chosen such that the force constants can be set to zero in the region III ($r>r_{outer}$).
As a result, the force constants in the region II ($r_{inner}\leq r\leq r_{outer}$) are responsible for effectively describing the 
remaining contribution and are going to be partitioned. 
Generally speaking, the suitable values of $r_{inner}$ and $r_{outer}$ depend on the studied system.
Considering that the supercell boundary should be included in the region II, 
the inscribed and circumscribed spheres centered at the supercell center can be chosen for the inner and outer spheres, respectively.
In this way, the radii of the two spheres, $r_{inner}$ and $r_{outer}$, can be easily defined for an arbitrary supercell, for example, 
the $r_{inner}$ and $r_{outer}$ illustrated in Fig.~\ref{fig:model2} (a).
This simple choice will be applied to four systems in Sec.~\ref{sec:application}.
For the case where the atoms outside the proposed outer sphere are deemed important, the $r_{outer}$ should be set to a longer value
to include those atoms. In  Fig.~\ref{fig:model2} (b), such a different choice is illustrated based on the physical consideration.
Note that all the atoms in a supercell can be translationally shifted to the supercell center so the same $r_{inner}$ and $r_{outer}$ 
can be adopted for all the atoms. 

Once $r_{inner}$ and $r_{outer}$ are determined, the weight of force constants among the periodic atoms in the region II that satisfy the 
translational symmetry described by the supercell lattice vectors can be redistributed. To unambiguously redistribute the weight
is not easy even if the force constants are purely described by the dipole-dipole interactions. This is because different periodic atoms in
the region II share the same periodic images in the region III so that the exact form of dipole-dipole interaction 
cannot be unambiguously used to determine the effective weight. Nevertheless, it is still reasonable to assume that the weight for 
each periodic atom in the region II is inversely proportional to $r^d$, where $r$ denotes the radius of the corresponding circle, 
that is, the atom-reference atom distance. After applying the $r^{-d}$ weight for the atoms, the weight should be normalized to 1 to 
guarantee the unaffected solutions at the commensurate $q$ vectors. 

The suitable value of $d$ also depends on the studied system and should be treated as a tunable parameter. In fact, this single parameter 
does not need to reflect the decaying behavior of the force constants in the region I but is intended to recover the relative weight 
among the periodic atoms in the region II. Considering the rapidly decaying nature of the strength of force constants against distance, 
a value between 3 and 5, consistent with those described by the dipole-dipole interactions, could be a reasonable choice. 
For a larger value of $d$, whose effect mainly tends to redistribute the weight for the force constants associated with the 
shortest distance between the reference atom and the atoms in the region II, could also be a good choice. Therefore, the converged dispersion curves
with the increasing value of $d$, where the convergence means no prominent change, offer a way to estimate the suitable value of $d$, 
but such a value should avoid introducing new types of imaginary frequencies. 
The applicability of the proposed partition method and the suitable values of $d$ will be discussed for four systems in Sec.~\ref{sec:application}. 

In addition, it is worth mentioning that the partition method is compatible with the mixed-space method because there exist 
degrees of freedom in performing Eq.~\ref{eq:eq03} for a periodic system. 
Therefore, the same weight can be used to partition the non-analytic contribution.
In the DFPT calculations, it is common to use the exact dipole-dipole interactions, which do not need to follow the supercell periodicity, to separate 
the short-range contributions from the full contributions at the selected $q$-mesh and the real-space short-range force constants can then be 
obtained via the Fourier transform.\cite{PhysRevB.55.10355} 
However, it should be noted that while the separation 
is possible at each selected $q$ vector, the resultant short-range force constants can be modified owing to 
the degrees of freedom in choosing the real-space $R$ or reciprocal-space $q$ grids in performing the Fourier transform, where the issue
of partition becomes relevant. In the CSLD method, the separation of force constants into the 
short-range and the long-range dipole-dipole contributions is processed in real space and the Fourier transformation is also needed for 
constructing the dynamical matrices.\cite{PhysRevB.100.184309}
In this case, the physical properties of the fitted forces, such as the delivered symmetry and Hermiticity of the dynamical matrices,
are responsible for delivering accurate phonon dispersions.

\section{Applications}
\label{sec:application}

The applicability of the proposed partition method for the phonon dispersions in NaCl, PbTiO$_3$, monolayer CrI$_3$, and twisted bilayer graphene 
are studied here. All of the relaxed atomic forces in equilibrium are less than 0.0001 Ha/Bohr. 

\subsection{Phonon dispersion in NaCl}
\label{subsec:nacl}

NaCl, crystallizing in the rock salt structure, is adopted as the first example and the traditional partition method is expected to fail
in general due to the primitive-cell shape of the face-center-cubic structure (FCC).
The first-principles calculations were performed using OpenMX code,
where the generalized gradient approximation (GGA), norm-conserving pseudopotentials, and optimized pseudo-atomic basis functions 
were adopted.\cite{openmx,GGA,Theurich,Morrison,Ozaki}
Three, two, and one optimized radial functions were allocated for the $s$, $p$, and $d$ orbitals, respectively, for the Na atom with a
cutoff radius of 9 Bohr, denoted as Na9.0-s3p2d1. For the Cl atom, Cl7.0-s2p2d1 was used. A $12\times12\times12$ $k$-point sampling 
was adopted for the primitive unit cell. A cutoff energy of 400 Ha was used for the numerical integrations and for 
the solution of the Poisson equation. The structure, including the unit cell and atomic positions, is fully relaxed relying on the chosen
numerical grids without imposing the $Fm\bar{3}m$ symmetry. The relaxed lattice constant was found to be 5.72\AA.
The $6\times6\times6$ supercell, containing 432 atoms, was chosen for the phonon calculation, where the force constants were obtained 
by the adoption of a displacement of 0.05\AA.

\begin{figure}[tbp]
\includegraphics[width=0.98\columnwidth,clip=true,angle=0]{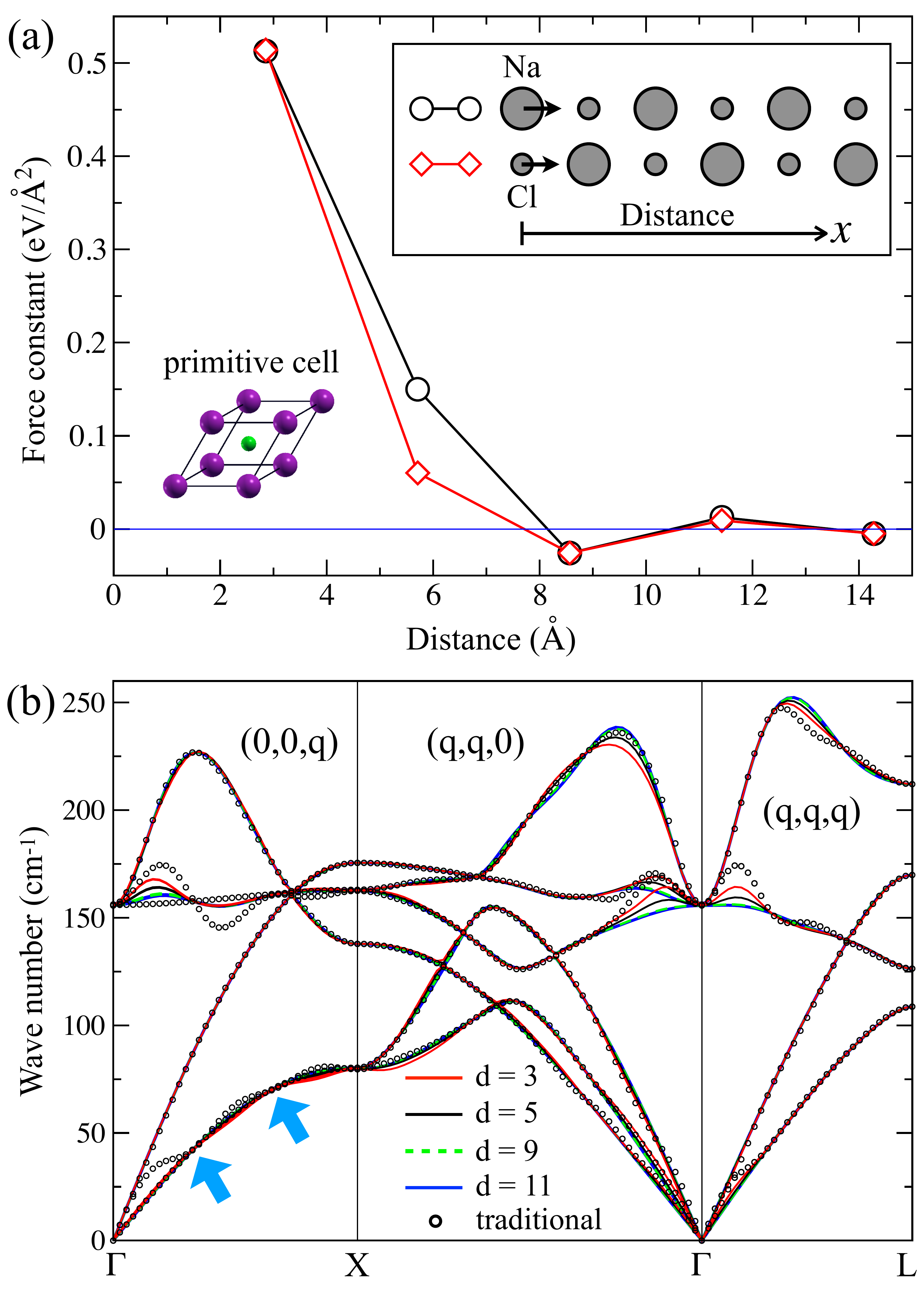}
\caption{(a) Force decay in NaCl along one dimension, calculated in the $6\times 6\times 6$ supercell containing 432 atoms 
using OpenMX code. Only the $x$-component forces of the $x$-direction neighboring atoms of Na or Cl atom that is displaced 
by~0.05\AA~along the $x$ direction are analyzed here. The force constants, obtained by dividing the forces by 0.05\AA, 
decay rapidly against the distance but still oscillate around zero at a long distance ($\sim12$\AA).
(b) Phonon dispersion of NaCl calculated using the $6\times 6\times 6$ supercell. The circles present the result using the traditional
partition method while the red, black, green, and blue curves present the results using the proposed partition method with $d=$
3, 5, 9, and 11, respectively. The arrows indicate two degenerate frequencies at the commensurate $q$ points between $\Gamma$ and X. 
LO-TO splitting is not considered here.
}
\label{fig:NaCl}
\end{figure}

After displacing one of the Na or Cl atoms towards its first neighbor in the $6\times6\times6$ supercell, 
say, along the $x$ direction, the experienced forces of 
the neighboring atoms along the same direction are expected to decay with the increasing interatomic distance. 
As shown in Fig.~\ref{fig:NaCl} (a), the strength of the forces drops significantly beyond $\sim6$\AA. However, the forces 
are not converged to zero but oscillate around zero even at a long distance of $\sim12$\AA, which suggests that a larger supercell
and denser real-space grids in describing the charge density and atomic basis functions are needed.
In Fig.~\ref{fig:NaCl} (b), without including LO-TO splitting, the phonon dispersion obtained from the traditional partition method 
is presented by the circles. Several expected degeneracies 
in NaCl\cite{PhysRev.178.1496,MARCONDES201811,PhysRevB.100.184309} are lifted, for example, 
the two transverse acoustic branches along the path between $\Gamma$ and X are not degenerate 
except those at the commensurate $q$ points indicated by the arrows. 
The unaffected degeneracies at $\Gamma$, X, L, and the other commensurate $q$ points demonstrate the powerfulness 
of the supercell force-constant phonon calculations.

The lifted degeneracies suggest that the $6\times 6\times 6$ supercell containing 432 atoms is not large enough to diminish the effect
of the primitive-cell shape of FCC. To improve the phonon dispersion at the non-commensurate $q$ points, the proposed partition method 
can be applied. Although the decaying behavior of the force constants against the distance has been investigated along
the $x$ direction, the result may change for different combinations of forces and displacements.   
Without resorting to a more complicated form or adopting an average value of $d$, the suitable value can be investigated
by examining the dispersions with different values of $d$, which are shown in Fig.~\ref{fig:NaCl} (b). 
Expectedly, the frequencies at the commensurate $q$ vectors are unaffected. Without explicitly applying 
the rotational symmetry, the degeneracy for the two transverse acoustic branches between $\Gamma$ and $X$ are recovered with 
the increasing value of $d$, and the oscillating behavior at the $q$ points close to $\Gamma$ is diminished and averaged to 
degenerate bands. This oscillation can exist in a larger supercell calculation in reflecting the suppressed LO-TO splitting 
in the long-wavelength limit. According to the dispersions, $d=5$ is a good choice for correcting the lifted degeneracy. 
Otherwise, $d=9$ can be chosen since it gives a converged result in comparison with the $d=11$ result. Therefore, the convergence can 
be used as a guidance to select the value of $d$ since recovering the proper weight of the force constants associated with the 
shortest distance between the reference atom and the atoms in the region II is important, especially in the case where the 
forces decay rapidly at the supercell boundary. For a small supercell, like the primitive unit cell itself, a large value of $d$ might 
give a result deviating from the exact solution owing to the expected longer-range distribution of the force constants, but 
the convergence can still be adopted for the guidance, which will be discussed below. 

To account for the LO-TO splitting in a polar system from first principles, 
the mixed-space approach or the linear response approach can be adopted. The mixed-space method proposes a consistent way to 
incorporate the non-analytic contribution with the interatomic force constants, where the Born effective charges 
and dielectric constants can be obtained from the linear response methods or 
via other calculations.\cite{PhysRevB.59.8551,PhysRevLett.89.157602,PhysRevB.47.1651}
The linear response method allows for the calculations of dielectric constants, 
Born effective charges, and the dynamical matrices using the same framework. 
For NaCl, the DFPT solutions can be obtained without heavy computations
and the result will be used to compare with those obtained from the supercell force-constant calculations. 

\begin{figure}[tbp]
\includegraphics[width=1.00\columnwidth,clip=true,angle=0]{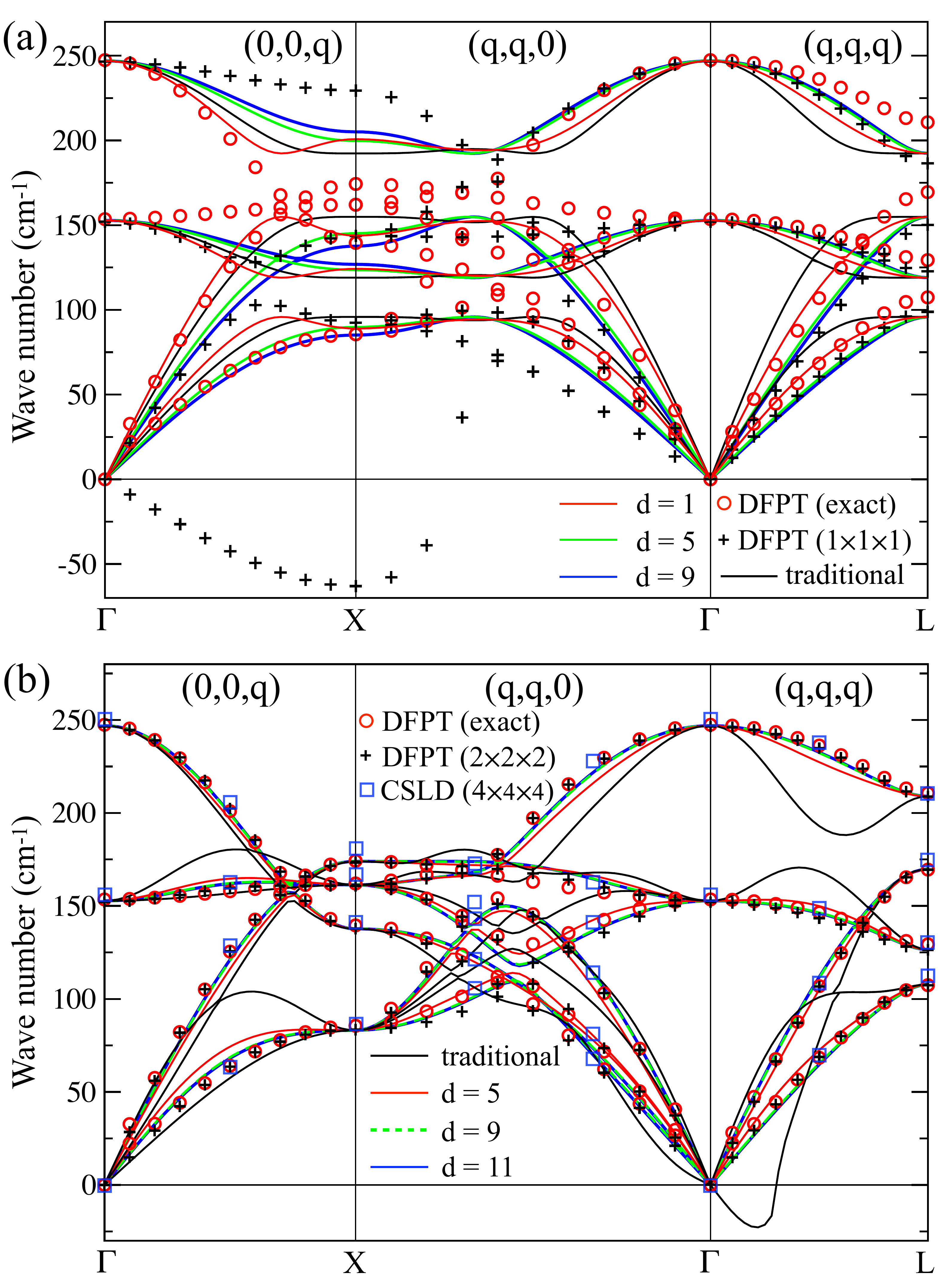}
\caption{Phonon dispersion of NaCl calculated using ABINIT code. 
The DFPT dispersions obtained from the Fourier interpolation with (a) $1\times1\times1$ and (b) $2\times2\times2$ $q$-meshes
are presented by the plus sign. 
The dispersions obtained from (a) $1\times 1\times 1$ and (b) $2\times 2\times 2$ supercell force-constant calculations using the 
traditional partition method are presented by the black curves while the red, green, and blue curves present
the results using the proposed partition method with different values of $d$ (see the legends). 
The DFPT dispersion directly calculated for the $q$ vectors along the path is presented by the red circles (exact). 
The dispersion obtained from the compressive sensing lattice dynamics (CSLD) calculations using the $4\times 4\times 4$ conventional 
cell\cite{PhysRevB.100.184309} is depicted by the squares.
}
\label{fig:linear}
\end{figure}

For consistency, the DFPT calculations\cite{PhysRevB.55.10337,PhysRevB.55.10355} together with
the $1\times1\times1$ and $2\times2\times2$ supercell force-constant calculations were performed 
using ABINIT code,\cite{GONZE2016106}
where norm-conserving pseudopotentials,\cite{PhysRevB.88.085117} plane-wave basis functions, and GGA were adopted. 
The cut-off energy of 120 Ha and the $12\times12\times12$ $k$-point sampling were adopted. 
The structure was fully relaxed under $Fm\bar{3}m$ space group symmetry. 
The relaxed lattice constant was found to be 5.69\AA.
The DFPT solutions directly calculated at the selected $q$ points are presented by the red circles in Fig.~\ref{fig:linear}, 
which can be treated as the exact solutions. The DFPT dispersions obtained from the Fourier interpolation with 
both $1\times1\times1$ and $2\times2\times2$ $q$-meshes, denoted as DFPT ($1\times1\times1$) and ($2\times2\times2$), respectively, 
are presented by the plus sign. In the supercell force-constant calculations, 
the displacement of~0.05\AA~was adopted. The dielectric constant, 2.487, and the Born effective charges, Na: 1.106 and Cl: -1.106, 
obtained from DFPT were used for constructing the non-analytic term of dynamical matrix.

The result of $1\times1\times1$ supercell force-constant calculations is shown in Fig.~\ref{fig:linear} (a).
In this primitive-cell calculation involving only two symmetry-related atoms, the dispersion using the 
traditional partition method can capture the overall behavior of the exact dispersion except there are no degrees 
of freedom in splitting the acoustic branches into three branches along the $(q,q,0)$ path. The dispersions 
using our proposed partition method with different values of $d$ indicate that increasing the value of $d$ does not 
always bring the dispersion closer to the exact solution. For example, $d=1$ can give a better result than $d=9$ for some branches. 
But the result with $d=9$, which is converged to the result with $d=11$ (not shown), still provides good estimation of the 
phonon dispersion. Although the detailed description for the partition method implemented in the ABINIT code is unavailable in the literature,
the DFPT ($1\times1\times1$) calculated using the ABINIT code is presented for reference. 
It is found that such an implementation is able to recover the correct number of splitting. 
However, imaginary frequencies have also been introduced in the dispersion. 

The results of $2\times2\times2$ supercell force-constant calculations and DFPT ($2\times2\times2$) are presented in Fig.~\ref{fig:linear} (b).
The result of DFPT ($2\times2\times2$) is now in good agreement with the exact DFPT solution.
For the supercell force-constant calculations, since the supercell size is not large, 
the directly calculated force constants must encounter a highly asymmetric 
distribution due to the shape of the supercell. The phonon dispersion using the traditional partition method presented by 
the black curves in Fig.~\ref{fig:linear} (b) reflects such effects, for example, the degeneracy for the transverse 
acoustic branches between $\Gamma$ and $X$ has been lifted significantly.
More seriously, imaginary frequencies are found along the $\Gamma$-L path.
The red, green, and blue curves in Fig.~\ref{fig:linear} (b) present the results using our proposed partition 
method with $d=$ 5, 9, and 11, respectively. Similar to the $6\times6\times6$ supercell calculations, 
the phonon dispersion with $d=9$ is converged to that with $d=11$ and is in good agreement with both DFPT calculations.
The result with the properly partitioned force constants is also in good agreement with the CSLD result,\cite{PhysRevB.100.184309}  
as shown in Fig.~\ref{fig:linear} (b).

As shown in Fig.~\ref{fig:linear}, the form, $r^{-d}$, can provide good estimation of phonon
dispersion in both the $1\times1\times1$ and $2\times2\times2$ supercell force-constant calculations.
The results suggest that the value of $d$ giving rise to the result closer to the
exact solution depends on the supercell size. Without having a good reference for the phonon dispersion, the convergence of 
dispersion with the increasing value of $d$ can be adopted as the guidance for determining the suitable value of $d$ 
for a reasonable interpolation. However,
the convergence needs to be examined with caution by gradually increasing the value of $d$. An extremely large value of $d$ 
could easily give incorrect weight for tiny numerical noise, for example, two periodic atoms having almost the same distance to the reference atom 
can be assigned with very different weight. As discussed earlier, the value of $d$ should avoid introducing
new types of imaginary frequencies, where maximizing the weight for the atoms in the region II with the shortest distance to the reference atom 
becomes inappropriate.
Finally, we emphasize that the proper partition is important for improving not only the 
phonon dispersion but also the vibrational modes since the highly asymmetric distribution of force constants can give unphysical 
vibrations.

\subsection{Phonon dispersion in PbTiO$_3$}

\begin{figure}[tbp]
\includegraphics[width=1.00\columnwidth,clip=true,angle=0]{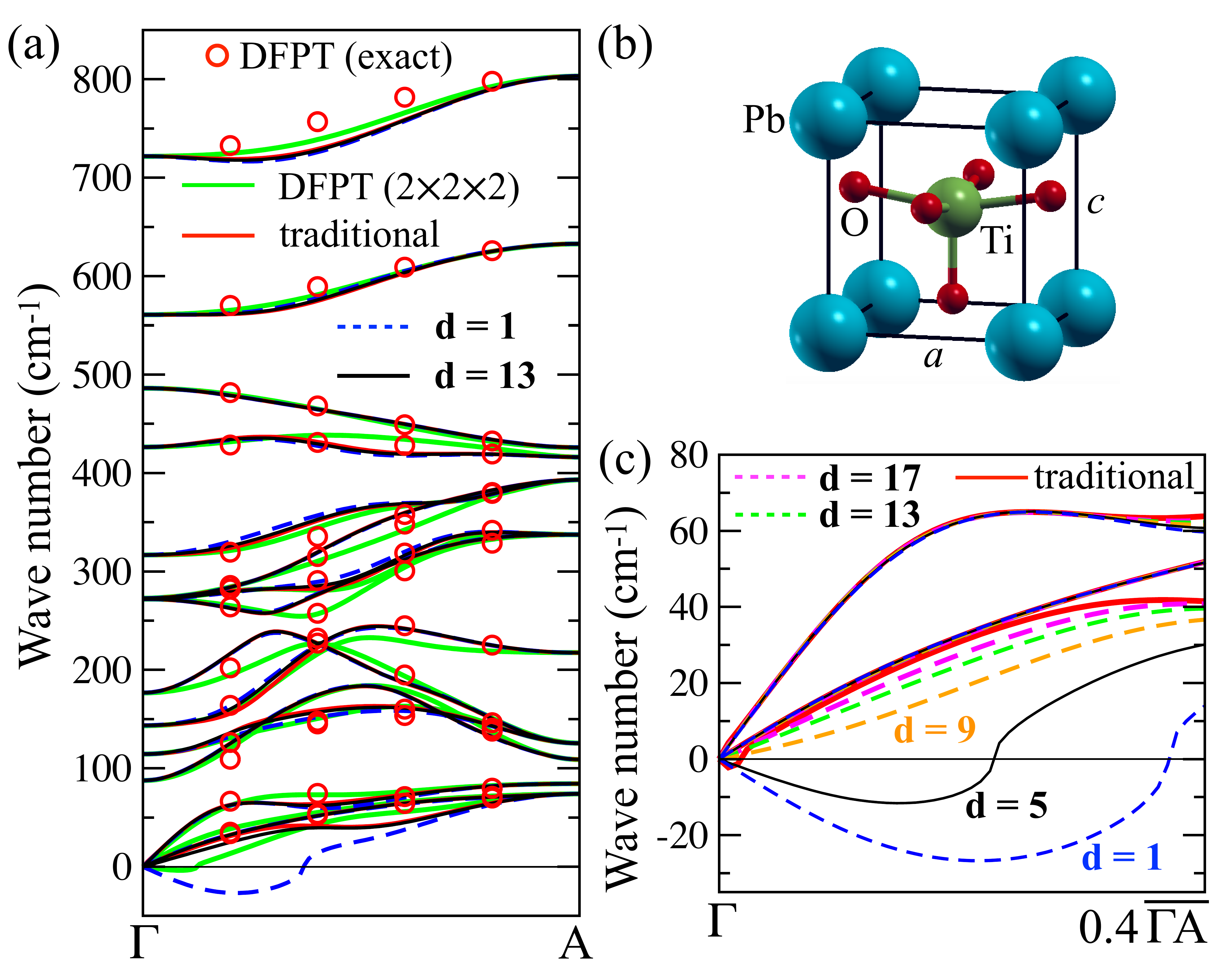}
\caption{(a) Phonon dispersion in PbTiO$_3$ between $\Gamma$ and A(0.5, 0.5, 0.5) calculated using ABINIT code.
The exact DFPT dispersion is presented by the red circles.
The DFPT ($2\times2\times2$) result is presented by the green curves.
The dispersion obtained from the $2\times2\times2$ supercell calculations using the
traditional partition method is presented by the red curves while the blue and black curves present
the results using the proposed partition method with $d$=1 and 13, respectively.
(b) Structure of PbTiO$_3$. (c) Phonon dispersions obtained using the traditional partition method and
the proposed partition method with different values of $d$ (see the legends) around $\Gamma$.
}
\label{fig:pbtio3}
\end{figure}

The first-principles calculations of PbTiO$_3$ were performed using ABINIT code within GGA. 
The cut-off energy of 120 Ha, the $6\times6\times6$ $k$-point sampling, 
and the lattice constants of $a=$ 3.902\AA~and $c=$ 4.156\AA~were adopted.
The structure is shown in Fig.~\ref{fig:pbtio3} (b).
For the phonon calculations, we focus on the $2\times2\times2$ supercell force-constant calculations
and the DFPT dispersion obtained from the Fourier interpolation with the $2\times2\times2$ $q$-mesh. 
The displacement of~$0.01$\AA~was adopted for the supercell force-constant calculations.
The $P4mm$ space group symmetry was imposed in the DFPT calculations.
The DFPT dispersion calculated at the selected $q$ points, which is considered as the exact solution, is presented by the red circles
in Fig.~\ref{fig:pbtio3} (a).
The Fourier-interpolated dispersion, denoted as DFPT ($2\times2\times2$), is presented by the green curves.
The dielectric constants and the Born effective charges needed for describing the LO-TO splitting in the supercell force-constant method 
were obtained from the DFPT calculations for consistency.

Overall, both dispersions obtained from the DFPT ($2\times2\times2$) calculations and the 
$2\times2\times2$ supercell force-constant calculations using the
traditional partition method are in good agreement with the exact DFPT solution 
except the undesired imaginary frequencies in the acoustic branch between $\Gamma$ and A, 
as shown in Fig.~\ref{fig:pbtio3} (a).
The acoustic frequencies around $\Gamma$ can be altered by multiple sources, such as 
the numerical noise induced by the calculation precision,\cite{Lucas}
the imposed acoustic sum rule for reaching zero frequencies at $\Gamma$,
and lacking the rotational symmetry or Hermiticity in the dynamical matrices.\cite{PhysRevB.100.184309}
The incorrect description of higher-order multipolar interactions in the 
piezoelectric materials could also lead to imaginary frequencies.\cite{Stengel}
Here, we focus on the effect of partitioning force constants in the region II on the acoustic frequencies around $\Gamma$.
The force constants associated with the moderate-size $2\times2\times2$ supercell allow us to demonstrate the usefulness of 
the proposed partition method. 

The phonon dispersions obtained using the proposed partition method with the values of $d=1$, 5, 9, 13, and 17 are 
presented in Fig.~\ref{fig:pbtio3} (c) for the path between $\Gamma$ and 2/5 of $\Gamma$ to A.
In this scale, the traditional partition method is observed to deliver imaginary frequencies 
and the proposed partition method with $d=1$ gives a band having more imaginary-frequency modes. 
However, with the increasing value of $d$, a trend of disappearance of imaginary-frequency modes can be clearly identified. 
By switching the value of $d$ from 5 to 9, the imaginary-frequency modes disappear, similar to the case experiencing 
a phase transition. The comparison between the results with $d=13$ and $d=17$ indicates the slow convergence 
of the lowest-frequency band against the value of $d$. The dispersions with different values of $d$ 
demonstrate the sensitivity of the acoustic frequencies to the redistribution of the weight among the atoms in the 
region II, and it is expected that the frequencies can be easily altered by the numerical noise introduced to
the calculations. In the present case, the proposed partition method can systematically recover the proper weight
among the atoms in the region II. In Sec.~\ref{subsec:graphene}, we will show that a larger supercell is needed
to more accurately describe the weight in the original region II where the simple form, $r^{-d}$, cannot further improve
the dispersion.

\subsection{Phonon dispersion in monolayer CrI$_3$}

\begin{figure}[tbp]
\includegraphics[width=1.00\columnwidth,clip=true,angle=0]{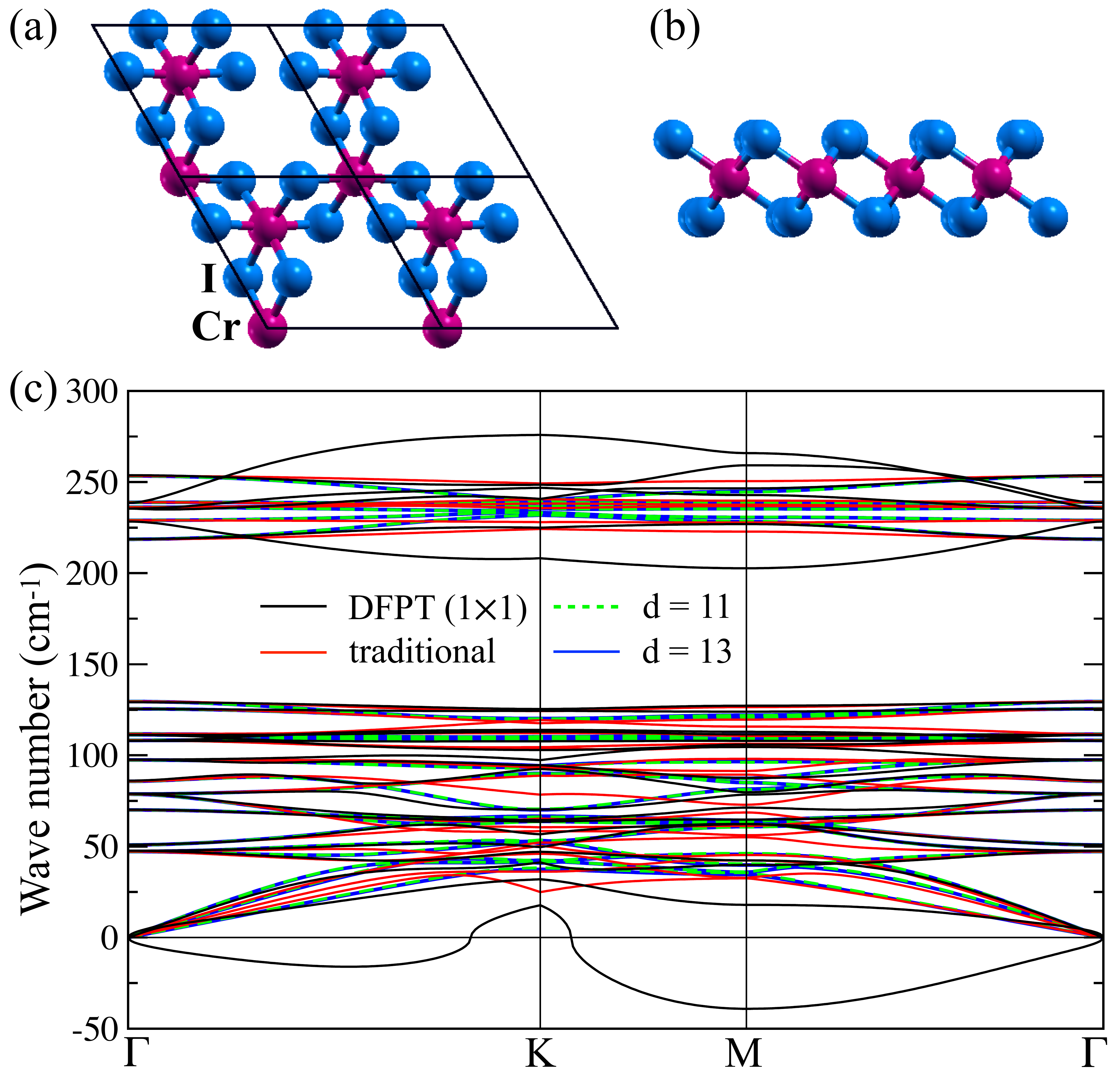}
\caption{
The (a) top and (b) side views of antiferromagnetic monolayer CrI$_3$. 
(c) The DFPT ($1\times1$) result calculated using ABINIT code is presented by the black curves.
The dispersion obtained from the $1\times1$ supercell calculations using the
traditional partition method are presented by the red curves while the green and blue curves present
the results using the proposed partition method with $d$=11 and 13, respectively.
}
\label{fig:cri3}
\end{figure}

\begin{figure}[tbp]
\includegraphics[width=1.00\columnwidth,clip=true,angle=0]{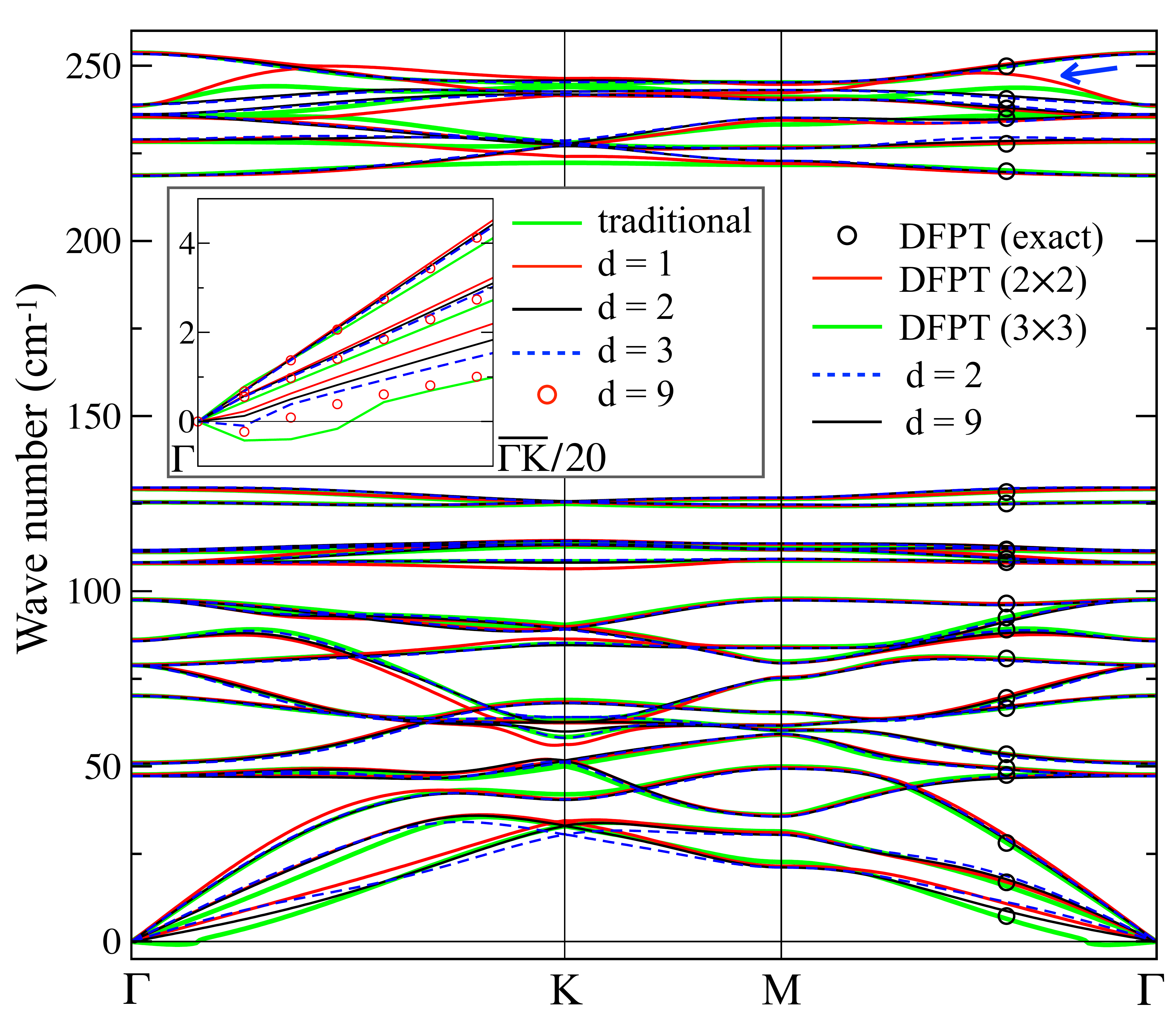}
\caption{
The DFPT dispersions from Fourier interpolation with $2\times2$ and $3\times3$ $q$-meshes using ABINIT code 
are presented by the red and green curves, respectively. The dispersions from the $2\times2$ supercell force-constant 
calculations using the proposed partition method with $d=2$ and 9 are presented by the blue and black curves, respectively.
The DFPT result directly calculated for the selected $q$ vectors is presented by the circles (exact).
The interpolated bands where the proposed partition method gives better agreement with the exact solution than the Fourier interpolation
implemented by ABINIT code are indicated by the arrow. In the inset, the acoustic branches using the traditional partition method
and different values of $d$ are presented between $\Gamma$ and 1/20 of $\Gamma$ to K path (see the legends).
}
\label{fig:cri3ano}
\end{figure}

The first-principles calculations of monolayer CrI$_3$ were performed using ABINIT code within 
the local density approximation (LDA).
The cut-off energy of 40 Ha, the $6\times6$ $k$-point sampling,
and the hexagonal lattice constant of $a=6.859$\AA~were adopted for the calculations.
The atomic positions were relaxed without imposing the space group symmetry and
the CrI$_3$ layers were separated by $\sim20$\AA~to avoid the interlayer interactions. 
The primitive unit cell containing two Cr atoms forming the antiferromagnetic order is presented in 
Figs.~\ref{fig:cri3} (a) and (b). 
For the phonon calculations, the $1\times1$ and $2\times2$ supercell force-constant calculations
were performed with the displacement of~$0.1$\AA. 
The DFPT dispersions obtained from the Fourier interpolations with the $1\times1$, $2\times2$, and $3\times3$ $q$-meshes, 
denoted as DFPT ($1\times1$), ($2\times2$), and ($3\times3$) were presented in Fig.~\ref{fig:cri3} (c)
and Fig.~\ref{fig:cri3ano}, respectively.
The dielectric constants and the Born effective charges needed for describing the LO-TO splitting in the supercell force-constant method 
were obtained from the DFPT calculations for consistency.

The dispersions obtained from the $1\times1$ supercell force-constant calculations using the traditional partition method and the 
proposed partition method with $d=11$ and $d=13$ are shown in Fig.~\ref{fig:cri3} (c).
The $d=11$ result, which is converged to the $d=13$ result,
does not reveal a significant difference from the result using the traditional partition method except the band around the lowest frequency at K. 
Even though the calculations were performed using the primitive unit cell, the result with $d=11$ can well capture the overall 
dispersions obtained from the DFPT ($2\times2$) and ($3\times3$) calculations. 
By contrast, the DFPT ($1\times1$) result, 
which gives the consistent frequencies at $\Gamma$, reveals a band having the imaginary frequencies that cannot be overlooked.
As shown in Fig.~\ref{fig:cri3ano}, most of the imaginary frequencies become positive in the DFPT ($2\times2$) and ($3\times3$) calculations. 
More specifically, the imaginary frequencies exist only at the $q$ points closer to $\Gamma$ in the DFPT ($3\times3$) calculations. 
In the DFPT ($2\times2$) result, the imaginary frequencies having the similar unphysical shape around $\Gamma$ can be found without 
applying the acoustic sum rule (not shown). Such imaginary frequencies around $\Gamma$ have also been reported in Ref.~\onlinecite{Lucas}. 
As shown in the inset of Fig.~\ref{fig:cri3ano}, the same type of imaginary frequencies also exist in the 
$2\times2$ supercell force-constant calculations using the traditional partition method.

Regardless of the imaginary frequencies around $\Gamma$, 
both dispersions obtained from the $2\times2$ supercell force-constant calculations using the 
proposed partition method with $d=2$ and $9$, where
the $d=9$ result is converged to the $d=11$ one (not shown), are in good agreement with the 
DFPT ($2\times2$) and ($3\times3$) results. Concerning the imaginary frequencies around $\Gamma$, 
the result using $d=9$ is found to possess an imaginary frequency at the chosen $q$ grids shown
in the inset of Fig.~\ref{fig:cri3ano}. Since this imaginary-frequency band is much shallower 
than that obtained from the DFPT ($3\times3$) calculations, the imaginary frequencies could come from the numerical noise.
Nevertheless, the effect of different values of $d$ can still be explored. The results with $d=1$ and 2 are found to carry 
no imaginary frequencies. For $d=3$, the frequency becomes imaginary 
and remains imaginary up to $d=9$. Since the increasing value of $d$ should avoid introducing the imaginary frequencies,
$d=2$ can be chosen to keep the band on the positive-frequency side.
Finally, we address the effect of the proposed partition method on the high-frequency band. 
In comparison with the exact solution presented by the circles in Fig.~\ref{fig:cri3ano}, 
the proposed partition method with $d=2$ is found to give better agreement in describing the high frequencies 
than the DFPT ($2\times2$) and ($3\times3$) calculations, as pointed out by the arrow shown in Fig.~\ref{fig:cri3ano}.

\subsection{Phonon dispersion in twisted bilayer graphene}
\label{subsec:graphene}

\begin{figure}[tbp]
\includegraphics[width=1.00\columnwidth,clip=true,angle=0]{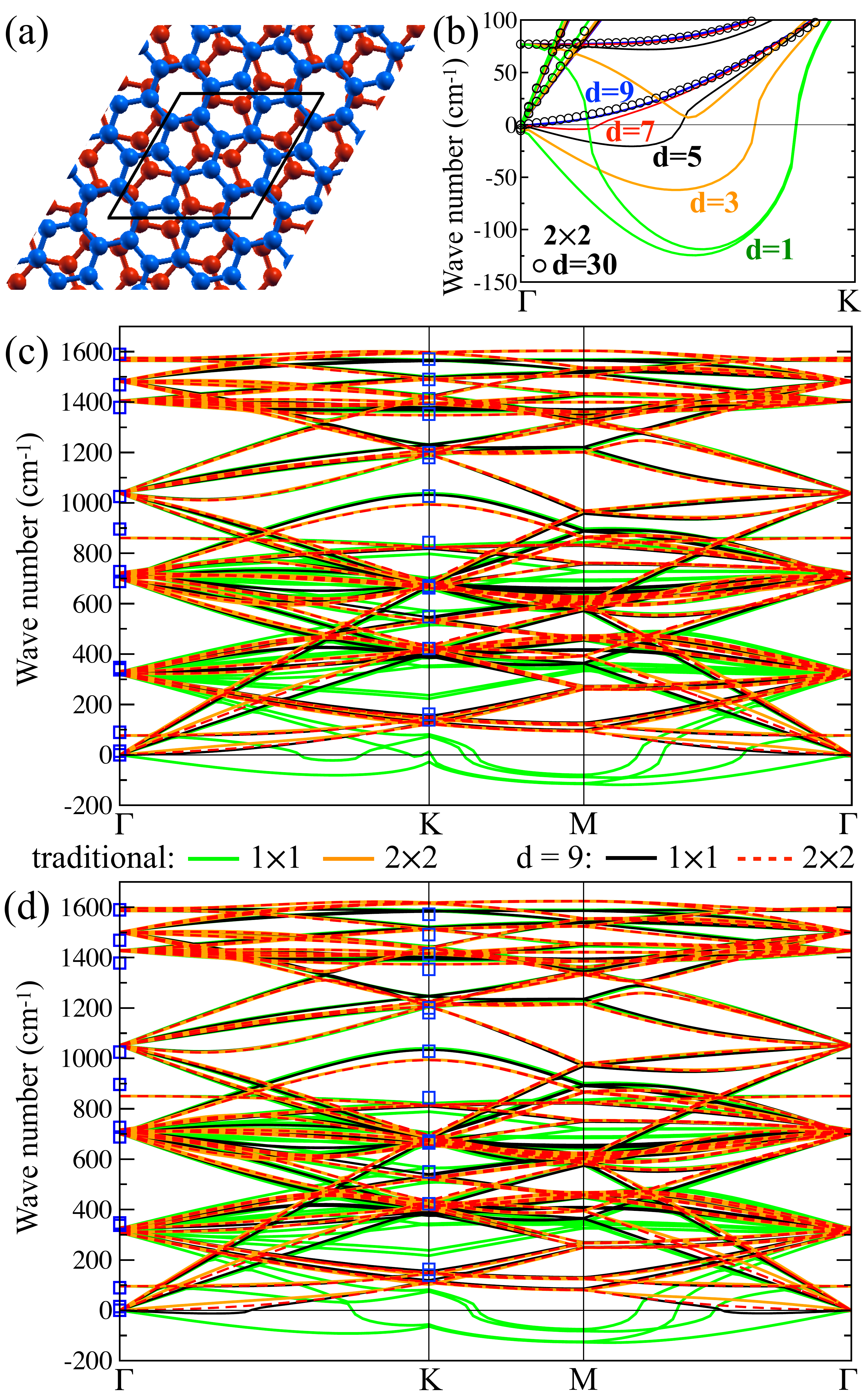}
\caption{(a) Structure of twisted bilayer graphene with a rotational angle of 21.8$^\circ$.
(b) Dispersions obtained from the $1\times1$ supercell force-constant calculations with different values of $d$ (see the legends)
within LDA using OpenMX code. The LDA dispersion from the $2\times2$ supercell calculations with $d=30$ is presented by the circles.
The results using the traditional partition method and the proposed partition method with $d=9$
obtained from the $1\times1$ and $2\times2$ supercell calculations within
(c) LDA and (d) DFT-D2 are presented by different curves (see the legends). 
The frequencies at $\Gamma$ and K obtained from Ref.~\onlinecite{PhysRevB.88.035428} based on 
the Born–von Karman model and the Lennard-Jones potential are presented by the squares.
}
\label{fig:tblg}
\end{figure}

The phonon dispersion of twisted bilayer graphene (T-BLG) has been studied using the Born–von Karman model 
and the Lennard-Jones potential for the intralayer and interlayer interactions, respectively.\cite{PhysRevB.88.035428} 
Assisted by the first-principles frozen-phonon method, the breathing and shear vibrational modes in T-BLG have also been  
investigated.\cite{SONG20192628} In particular, the shear modes carrying almost zero frequencies in T-BLG, 
which are difficult to be accurately analyzed by the supercell force-constant and linear response approaches, 
can be studied by calculating the total energy against the shear vibration via displacing the graphene monolayer as a rigid body. 
Different from the direct calculation of dynamical matrix, the phonon frequencies are also reported using molecular dynamics 
simulations.\cite{RAMNANI2017302}
Here, the first-principles phonon dispersion for T-BLG with a rotational angle of 21.8$^\circ$, as shown in Fig.~\ref{fig:tblg} (a), 
is presented using the proposed partition method based on the LDA 
and the corrected GGA with the van der Waals interactions (DFT-D2).\cite{grimme2006semiempirical}

The first-principles calculations were performed using OpenMX code. C7.0-s2p2d1 was chosen for the basis functions. 
A cutoff energy of 400 Ha was used for the numerical integrations and for the solution of the Poisson equation. 
A $12\times12$ $k$-point sampling was adopted for the primitive unit cell. 
In general, T-BLG could contain a large amount of atoms in the primitive unit cell. 
For the $\theta=21.8^\circ$ case, the unit cell contains 28 atoms, which can be studied using atomic basis functions 
without heavy computations.
The experimental in-plane lattice constant ($a=6.5190$\AA)\cite{SONG20192628} was used and 
$c=20$\AA~was adopted to avoid the interactions between T-BLG images. 
The relaxed interlayer distances by LDA and DFT-D2 were found to be 3.4\AA~and 3.3\AA, respectively. 
For the phonon calculations, the $1\times1$ and $2\times2$ supercell force-constant calculations
were performed with the displacement of~$0.01$\AA. 

Within GGA, the interlayer distance is found to be unreasonably long and therefore cannot well describe the 
interlayer breathing mode. The frequencies obtained from the LDA and DFT-D2 calculations are 77.04 $cm^{-1}$ and 95.53 $cm^{-1}$, respectively.
The reported frequencies for the interlayer breathing mode are 89.5 $cm^{-1}$,\cite{PhysRevB.88.035428}
86.5 $cm^{-1}$,\cite{SONG20192628} and 94.56 $cm^{-1}$.\cite{RAMNANI2017302}
This suggests that to describe the interlayer interactions, the inclusion of the van der Waals interactions, 
as implemented in DFT-D2, can deliver the frequency that is consistent with experiments.\cite{RAMNANI2017302} 
In the $1\times1$ supercell force-constant calculations, the results with $d=9$ are found to be converged with the $d=11$ ones in 
both the LDA and DFT-D2 calculations. The LDA dispersions near $\Gamma$ with $d=1,$ 3, 5, 7, and 9 can be found in Fig.~\ref{fig:tblg} (b).
The phonon dispersions using the traditional partition method and the proposed partition method with $d=9$ are presented 
in Figs.~\ref{fig:tblg} (c) and (d) for the LDA and DFT-D2 functionals, respectively, where
the results are in good agreement with the phonon dispersion calculated using the Born–von Karman model and the Lennard-Jones potential.\cite{PhysRevB.88.035428} 
Overall, the effect of the proposed partition method is found to improve the description of the low-frequency bands
in comparison with those obtained from the more accurate $2\times2$ supercell force-constant calculations, for example, 
turning the imaginary-frequency bands obtained from the traditional partition  method to the positive-frequency bands, which demonstrates the
usefulness of the proposed partition method.

As shown by the black curves in Fig.~\ref{fig:tblg} (d), the proposed partition method cannot fully eliminate the imaginary frequencies around $\Gamma$
in the DFT-D2 calculations using the primitive cell. The result has been improved in the $2\times2$ supercell force-constant calculations, where the dispersion  
using either the traditional partition method or the proposed partition method with $d=9$ possesses no imaginary frequencies except the nearly vanishing imaginary
frequencies of the shear modes at $\Gamma$. 
Note that the imaginary shear-mode frequencies cannot be eliminated by any partition method but computational parameters because $q=(0,0,0)$ is 
an exact $q$ point. 
The improvement found in the $2\times2$ supercell is understandable since more accurate weight of the force constants
in the region II of the primitive cell can be obtained. 
In the $2\times2$ supercell calculations within DFT-D2, a larger value of $d$, say $d=11$, can introduce imaginary frequencies, 
showing the sensitivity of the dispersion to the redistribution of the weight of the force constants 
in the region II of the $2\times2$ supercell. 

On the other hand, the LDA dispersion can reach good convergence without introducing
imaginary frequencies. The distribution of the force constants in the region II described by the LDA are robust against larger values of $d$ in 
both the $1\times1$ and $2\times2$ supercell calculations. The LDA result with $d=30$ is presented in Fig.~\ref{fig:tblg} (b) for reference.
Although the frequency of the breathing mode is well described by DFT-D2,
we note that not all of the frequencies obtained by DFT-D2 are closer to the reported frequencies. 
For the modes with the frequency of 896.36 $cm^{-1}$ at $\Gamma$,\cite{PhysRevB.88.035428} 
both LDA and DFT-D2 give lower frequencies, namely, 860.48 $cm^{-1}$ and 850.26 $cm^{-1}$, respectively. 
The calculated LDA and DFT-D2 phonon dispersions together with the reported frequencies using different approaches\cite{PhysRevB.88.035428,RAMNANI2017302,SONG20192628} 
can be used for the interpretation of experimental data in the future.

\section{Summary}
\label{sec:summary}

We have addressed the importance of proper partition of force constants in the supercell force-constant phonon calculations. 
Without a proper partition, the resultant phonon frequencies and vibrations could be unphysical. 
Using simple model structures, we have shown that applying symmetry operations 
can improve the phonon properties but should be used with caution; otherwise, all of the phonon properties, including
those at the commensurate $q$ points, can be affected. While there are abundant ways to partition the force constants 
as can be confirmed by the discussed formulation, we propose a simple way to partition the force constants based only 
on the interatomic distance and translational symmetry. The proposed partition method is compatible with the 
mixed-space method for delivering accurate LO-TO splitting.
The phonon dispersions in NaCl, PbTiO$_3$, and monolayer CrI$_3$ with the properly partitioned force constants are demonstrated to be 
consistent with those obtained by the linear response method.
The proposed partition method can improve not only the low-frequency dispersion but also the high-frequency one, as demonstrated in the case of CrI$_3$. 
Finally, we present the phonon dispersion of twisted bilayer graphene with a rotational angle of 21.8$^\circ$ calculated from first principles. 
The obtained phonon dispersions are consistent with those calculated using the Born–von Karman model and the Lennard-Jones potential. 
The partition method can also be guided by symmetry operations and/or force-constant models as long as the sum rule (Eq.~\ref{eq:eq07}) is satisfied. 

\begin{acknowledgments}
Chi-Cheng Lee acknowledges the Ministry of Science and Technology of Taiwan for financial support under contract No. MOST 108-2112-M-032-010-MY2.
Hung-Chung Hsueh acknowledges the Ministry of Science and Technology of Taiwan for financial support under contract No. MOST 107-2112-M-032-009-MY3.
\end{acknowledgments}

\bibliography{refs}

\begin{thebibliography}{50}%
\makeatletter
\providecommand \@ifxundefined [1]{%
 \@ifx{#1\undefined}
}%
\providecommand \@ifnum [1]{%
 \ifnum #1\expandafter \@firstoftwo
 \else \expandafter \@secondoftwo
 \fi
}%
\providecommand \@ifx [1]{%
 \ifx #1\expandafter \@firstoftwo
 \else \expandafter \@secondoftwo
 \fi
}%
\providecommand \natexlab [1]{#1}%
\providecommand \enquote  [1]{``#1''}%
\providecommand \bibnamefont  [1]{#1}%
\providecommand \bibfnamefont [1]{#1}%
\providecommand \citenamefont [1]{#1}%
\providecommand \href@noop [0]{\@secondoftwo}%
\providecommand \href [0]{\begingroup \@sanitize@url \@href}%
\providecommand \@href[1]{\@@startlink{#1}\@@href}%
\providecommand \@@href[1]{\endgroup#1\@@endlink}%
\providecommand \@sanitize@url [0]{\catcode `\\12\catcode `\$12\catcode
  `\&12\catcode `\#12\catcode `\^12\catcode `\_12\catcode `\%12\relax}%
\providecommand \@@startlink[1]{}%
\providecommand \@@endlink[0]{}%
\providecommand \url  [0]{\begingroup\@sanitize@url \@url }%
\providecommand \@url [1]{\endgroup\@href {#1}{\urlprefix }}%
\providecommand \urlprefix  [0]{URL }%
\providecommand \Eprint [0]{\href }%
\providecommand \doibase [0]{https://doi.org/}%
\providecommand \selectlanguage [0]{\@gobble}%
\providecommand \bibinfo  [0]{\@secondoftwo}%
\providecommand \bibfield  [0]{\@secondoftwo}%
\providecommand \translation [1]{[#1]}%
\providecommand \BibitemOpen [0]{}%
\providecommand \bibitemStop [0]{}%
\providecommand \bibitemNoStop [0]{.\EOS\space}%
\providecommand \EOS [0]{\spacefactor3000\relax}%
\providecommand \BibitemShut  [1]{\csname bibitem#1\endcsname}%
\let\auto@bib@innerbib\@empty
\bibitem [{\citenamefont {Hohenberg}\ and\ \citenamefont
  {Kohn}(1964)}]{Hohenberg}%
  \BibitemOpen
  \bibfield  {author} {\bibinfo {author} {\bibfnamefont {P.}~\bibnamefont
  {Hohenberg}}\ and\ \bibinfo {author} {\bibfnamefont {W.}~\bibnamefont
  {Kohn}},\ }\href {https://doi.org/10.1103/PhysRev.136.B864} {\bibfield
  {journal} {\bibinfo  {journal} {Phys. Rev.}\ }\textbf {\bibinfo {volume}
  {136}},\ \bibinfo {pages} {B864} (\bibinfo {year} {1964})}\BibitemShut
  {NoStop}%
\bibitem [{\citenamefont {Kohn}\ and\ \citenamefont {Sham}(1965)}]{Kohn}%
  \BibitemOpen
  \bibfield  {author} {\bibinfo {author} {\bibfnamefont {W.}~\bibnamefont
  {Kohn}}\ and\ \bibinfo {author} {\bibfnamefont {L.~J.}\ \bibnamefont
  {Sham}},\ }\href {https://doi.org/10.1103/PhysRev.140.A1133} {\bibfield
  {journal} {\bibinfo  {journal} {Phys. Rev.}\ }\textbf {\bibinfo {volume}
  {140}},\ \bibinfo {pages} {A1133} (\bibinfo {year} {1965})}\BibitemShut
  {NoStop}%
\bibitem [{\citenamefont {Baroni}\ \emph {et~al.}(2001)\citenamefont {Baroni},
  \citenamefont {de~Gironcoli}, \citenamefont {Dal~Corso},\ and\ \citenamefont
  {Giannozzi}}]{RevModPhys.73.515}%
  \BibitemOpen
  \bibfield  {author} {\bibinfo {author} {\bibfnamefont {S.}~\bibnamefont
  {Baroni}}, \bibinfo {author} {\bibfnamefont {S.}~\bibnamefont
  {de~Gironcoli}}, \bibinfo {author} {\bibfnamefont {A.}~\bibnamefont
  {Dal~Corso}},\ and\ \bibinfo {author} {\bibfnamefont {P.}~\bibnamefont
  {Giannozzi}},\ }\href {https://doi.org/10.1103/RevModPhys.73.515} {\bibfield
  {journal} {\bibinfo  {journal} {Rev. Mod. Phys.}\ }\textbf {\bibinfo {volume}
  {73}},\ \bibinfo {pages} {515} (\bibinfo {year} {2001})}\BibitemShut
  {NoStop}%
\bibitem [{\citenamefont {Wang}\ \emph {et~al.}(2016)\citenamefont {Wang},
  \citenamefont {Shang}, \citenamefont {Fang}, \citenamefont {Liu},\ and\
  \citenamefont {Chen}}]{Wang2016}%
  \BibitemOpen
  \bibfield  {author} {\bibinfo {author} {\bibfnamefont {Y.}~\bibnamefont
  {Wang}}, \bibinfo {author} {\bibfnamefont {S.-L.}\ \bibnamefont {Shang}},
  \bibinfo {author} {\bibfnamefont {H.}~\bibnamefont {Fang}}, \bibinfo {author}
  {\bibfnamefont {Z.-K.}\ \bibnamefont {Liu}},\ and\ \bibinfo {author}
  {\bibfnamefont {L.-Q.}\ \bibnamefont {Chen}},\ }\href
  {https://doi.org/10.1038/npjcompumats.2016.6} {\bibfield  {journal} {\bibinfo
   {journal} {npj Comput. Mater.}\ }\textbf {\bibinfo {volume} {2}},\ \bibinfo
  {pages} {16006} (\bibinfo {year} {2016})}\BibitemShut {NoStop}%
\bibitem [{\citenamefont {Petretto}\ \emph {et~al.}(2018)\citenamefont
  {Petretto}, \citenamefont {Dwaraknath}, \citenamefont {Miranda},
  \citenamefont {Winston}, \citenamefont {Giantomassi}, \citenamefont {van
  Setten}, \citenamefont {Gonze}, \citenamefont {Persson}, \citenamefont
  {Hautier},\ and\ \citenamefont {Rignanese}}]{Petretto}%
  \BibitemOpen
  \bibfield  {author} {\bibinfo {author} {\bibfnamefont {G.}~\bibnamefont
  {Petretto}}, \bibinfo {author} {\bibfnamefont {S.}~\bibnamefont
  {Dwaraknath}}, \bibinfo {author} {\bibfnamefont {H.~P.}\ \bibnamefont
  {Miranda}}, \bibinfo {author} {\bibfnamefont {D.}~\bibnamefont {Winston}},
  \bibinfo {author} {\bibfnamefont {M.}~\bibnamefont {Giantomassi}}, \bibinfo
  {author} {\bibfnamefont {M.~J.}\ \bibnamefont {van Setten}}, \bibinfo
  {author} {\bibfnamefont {X.}~\bibnamefont {Gonze}}, \bibinfo {author}
  {\bibfnamefont {K.~A.}\ \bibnamefont {Persson}}, \bibinfo {author}
  {\bibfnamefont {G.}~\bibnamefont {Hautier}},\ and\ \bibinfo {author}
  {\bibfnamefont {G.-M.}\ \bibnamefont {Rignanese}},\ }\href
  {https://doi.org/10.1038/sdata.2018.65} {\bibfield  {journal} {\bibinfo
  {journal} {Sci. Data}\ }\textbf {\bibinfo {volume} {5}},\ \bibinfo {pages}
  {180065} (\bibinfo {year} {2018})}\BibitemShut {NoStop}%
\bibitem [{\citenamefont {Wendel}\ and\ \citenamefont
  {Martin}(1978)}]{PhysRevLett.40.950}%
  \BibitemOpen
  \bibfield  {author} {\bibinfo {author} {\bibfnamefont {H.}~\bibnamefont
  {Wendel}}\ and\ \bibinfo {author} {\bibfnamefont {R.~M.}\ \bibnamefont
  {Martin}},\ }\href {https://doi.org/10.1103/PhysRevLett.40.950} {\bibfield
  {journal} {\bibinfo  {journal} {Phys. Rev. Lett.}\ }\textbf {\bibinfo
  {volume} {40}},\ \bibinfo {pages} {950} (\bibinfo {year} {1978})}\BibitemShut
  {NoStop}%
\bibitem [{\citenamefont {Cowley}(1968)}]{Cowley_1968}%
  \BibitemOpen
  \bibfield  {author} {\bibinfo {author} {\bibfnamefont {R.~A.}\ \bibnamefont
  {Cowley}},\ }\href {https://doi.org/10.1088/0034-4885/31/1/303} {\bibfield
  {journal} {\bibinfo  {journal} {Reports on Progress in Physics}\ }\textbf
  {\bibinfo {volume} {31}},\ \bibinfo {pages} {123} (\bibinfo {year}
  {1968})}\BibitemShut {NoStop}%
\bibitem [{\citenamefont {Zhou}\ \emph {et~al.}(2014)\citenamefont {Zhou},
  \citenamefont {Nielson}, \citenamefont {Xia},\ and\ \citenamefont
  {Ozoli\ifmmode \mbox{\c{n}}\else \c{n}\fi{}\ifmmode~\check{s}\else
  \v{s}\fi{}}}]{PhysRevLett.113.185501}%
  \BibitemOpen
  \bibfield  {author} {\bibinfo {author} {\bibfnamefont {F.}~\bibnamefont
  {Zhou}}, \bibinfo {author} {\bibfnamefont {W.}~\bibnamefont {Nielson}},
  \bibinfo {author} {\bibfnamefont {Y.}~\bibnamefont {Xia}},\ and\ \bibinfo
  {author} {\bibfnamefont {V.}~\bibnamefont {Ozoli\ifmmode \mbox{\c{n}}\else
  \c{n}\fi{}\ifmmode~\check{s}\else \v{s}\fi{}}},\ }\href
  {https://doi.org/10.1103/PhysRevLett.113.185501} {\bibfield  {journal}
  {\bibinfo  {journal} {Phys. Rev. Lett.}\ }\textbf {\bibinfo {volume} {113}},\
  \bibinfo {pages} {185501} (\bibinfo {year} {2014})}\BibitemShut {NoStop}%
\bibitem [{\citenamefont {Kresse}\ \emph {et~al.}(1995)\citenamefont {Kresse},
  \citenamefont {Furthm\"{u}ller},\ and\ \citenamefont {Hafner}}]{Kresse_1995}%
  \BibitemOpen
  \bibfield  {author} {\bibinfo {author} {\bibfnamefont {G.}~\bibnamefont
  {Kresse}}, \bibinfo {author} {\bibfnamefont {J.}~\bibnamefont
  {Furthm\"{u}ller}},\ and\ \bibinfo {author} {\bibfnamefont {J.}~\bibnamefont
  {Hafner}},\ }\href {https://doi.org/10.1209/0295-5075/32/9/005} {\bibfield
  {journal} {\bibinfo  {journal} {Europhysics Letters ({EPL})}\ }\textbf
  {\bibinfo {volume} {32}},\ \bibinfo {pages} {729} (\bibinfo {year}
  {1995})}\BibitemShut {NoStop}%
\bibitem [{\citenamefont {Ackland}\ \emph {et~al.}(1997)\citenamefont
  {Ackland}, \citenamefont {Warren},\ and\ \citenamefont
  {Clark}}]{Ackland_1997}%
  \BibitemOpen
  \bibfield  {author} {\bibinfo {author} {\bibfnamefont {G.~J.}\ \bibnamefont
  {Ackland}}, \bibinfo {author} {\bibfnamefont {M.~C.}\ \bibnamefont
  {Warren}},\ and\ \bibinfo {author} {\bibfnamefont {S.~J.}\ \bibnamefont
  {Clark}},\ }\href {https://doi.org/10.1088/0953-8984/9/37/017} {\bibfield
  {journal} {\bibinfo  {journal} {Journal of Physics: Condensed Matter}\
  }\textbf {\bibinfo {volume} {9}},\ \bibinfo {pages} {7861} (\bibinfo {year}
  {1997})}\BibitemShut {NoStop}%
\bibitem [{\citenamefont {Parlinski}\ \emph {et~al.}(1997)\citenamefont
  {Parlinski}, \citenamefont {Li},\ and\ \citenamefont
  {Kawazoe}}]{PhysRevLett.78.4063}%
  \BibitemOpen
  \bibfield  {author} {\bibinfo {author} {\bibfnamefont {K.}~\bibnamefont
  {Parlinski}}, \bibinfo {author} {\bibfnamefont {Z.~Q.}\ \bibnamefont {Li}},\
  and\ \bibinfo {author} {\bibfnamefont {Y.}~\bibnamefont {Kawazoe}},\ }\href
  {https://doi.org/10.1103/PhysRevLett.78.4063} {\bibfield  {journal} {\bibinfo
   {journal} {Phys. Rev. Lett.}\ }\textbf {\bibinfo {volume} {78}},\ \bibinfo
  {pages} {4063} (\bibinfo {year} {1997})}\BibitemShut {NoStop}%
\bibitem [{\citenamefont {Parlinski}(1999)}]{Parlinski}%
  \BibitemOpen
  \bibfield  {author} {\bibinfo {author} {\bibfnamefont {K.}~\bibnamefont
  {Parlinski}},\ }\href {https://doi.org/10.1063/1.59457} {\bibfield  {journal}
  {\bibinfo  {journal} {AIP Conference Proceedings}\ }\textbf {\bibinfo
  {volume} {479}},\ \bibinfo {pages} {121} (\bibinfo {year}
  {1999})}\BibitemShut {NoStop}%
\bibitem [{\citenamefont {Gonze}(1997)}]{PhysRevB.55.10337}%
  \BibitemOpen
  \bibfield  {author} {\bibinfo {author} {\bibfnamefont {X.}~\bibnamefont
  {Gonze}},\ }\href {https://doi.org/10.1103/PhysRevB.55.10337} {\bibfield
  {journal} {\bibinfo  {journal} {Phys. Rev. B}\ }\textbf {\bibinfo {volume}
  {55}},\ \bibinfo {pages} {10337} (\bibinfo {year} {1997})}\BibitemShut
  {NoStop}%
\bibitem [{\citenamefont {Gonze}\ and\ \citenamefont
  {Lee}(1997)}]{PhysRevB.55.10355}%
  \BibitemOpen
  \bibfield  {author} {\bibinfo {author} {\bibfnamefont {X.}~\bibnamefont
  {Gonze}}\ and\ \bibinfo {author} {\bibfnamefont {C.}~\bibnamefont {Lee}},\
  }\href {https://doi.org/10.1103/PhysRevB.55.10355} {\bibfield  {journal}
  {\bibinfo  {journal} {Phys. Rev. B}\ }\textbf {\bibinfo {volume} {55}},\
  \bibinfo {pages} {10355} (\bibinfo {year} {1997})}\BibitemShut {NoStop}%
\bibitem [{\citenamefont {Calzolari}\ and\ \citenamefont
  {Nardelli}(2013)}]{Arrigo}%
  \BibitemOpen
  \bibfield  {author} {\bibinfo {author} {\bibfnamefont {A.}~\bibnamefont
  {Calzolari}}\ and\ \bibinfo {author} {\bibfnamefont {M.~B.}\ \bibnamefont
  {Nardelli}},\ }\href {https://doi.org/10.1038/srep02999} {\bibfield
  {journal} {\bibinfo  {journal} {Sci Rep}\ }\textbf {\bibinfo {volume} {3}},\
  \bibinfo {pages} {2999} (\bibinfo {year} {2013})}\BibitemShut {NoStop}%
\bibitem [{\citenamefont {Shang}\ \emph {et~al.}(2017)\citenamefont {Shang},
  \citenamefont {Carbogno}, \citenamefont {Rinke},\ and\ \citenamefont
  {Scheffler}}]{SHANG201726}%
  \BibitemOpen
  \bibfield  {author} {\bibinfo {author} {\bibfnamefont {H.}~\bibnamefont
  {Shang}}, \bibinfo {author} {\bibfnamefont {C.}~\bibnamefont {Carbogno}},
  \bibinfo {author} {\bibfnamefont {P.}~\bibnamefont {Rinke}},\ and\ \bibinfo
  {author} {\bibfnamefont {M.}~\bibnamefont {Scheffler}},\ }\href
  {https://doi.org/https://doi.org/10.1016/j.cpc.2017.02.001} {\bibfield
  {journal} {\bibinfo  {journal} {Computer Physics Communications}\ }\textbf
  {\bibinfo {volume} {215}},\ \bibinfo {pages} {26 } (\bibinfo {year}
  {2017})}\BibitemShut {NoStop}%
\bibitem [{\citenamefont {Soler}\ \emph {et~al.}(2002)\citenamefont {Soler},
  \citenamefont {Artacho}, \citenamefont {Gale}, \citenamefont {Garc{\'i}a},
  \citenamefont {Junquera}, \citenamefont {Ordej{\'o}n},\ and\ \citenamefont
  {S{\'a}nchez-Portal}}]{0953-8984-14-11-302}%
  \BibitemOpen
  \bibfield  {author} {\bibinfo {author} {\bibfnamefont {J.~M.}\ \bibnamefont
  {Soler}}, \bibinfo {author} {\bibfnamefont {E.}~\bibnamefont {Artacho}},
  \bibinfo {author} {\bibfnamefont {J.~D.}\ \bibnamefont {Gale}}, \bibinfo
  {author} {\bibfnamefont {A.}~\bibnamefont {Garc{\'i}a}}, \bibinfo {author}
  {\bibfnamefont {J.}~\bibnamefont {Junquera}}, \bibinfo {author}
  {\bibfnamefont {P.}~\bibnamefont {Ordej{\'o}n}},\ and\ \bibinfo {author}
  {\bibfnamefont {D.}~\bibnamefont {S{\'a}nchez-Portal}},\ }\href
  {http://stacks.iop.org/0953-8984/14/i=11/a=302} {\bibfield  {journal}
  {\bibinfo  {journal} {J. Phys.: Condens. Matter}\ }\textbf {\bibinfo {volume}
  {14}},\ \bibinfo {pages} {2745} (\bibinfo {year} {2002})}\BibitemShut
  {NoStop}%
\bibitem [{\citenamefont {Gillan}\ \emph {et~al.}(2007)\citenamefont {Gillan},
  \citenamefont {Bowler}, \citenamefont {Torralba},\ and\ \citenamefont
  {Miyazaki}}]{GILLAN200714}%
  \BibitemOpen
  \bibfield  {author} {\bibinfo {author} {\bibfnamefont {M.}~\bibnamefont
  {Gillan}}, \bibinfo {author} {\bibfnamefont {D.}~\bibnamefont {Bowler}},
  \bibinfo {author} {\bibfnamefont {A.}~\bibnamefont {Torralba}},\ and\
  \bibinfo {author} {\bibfnamefont {T.}~\bibnamefont {Miyazaki}},\ }\href
  {https://doi.org/https://doi.org/10.1016/j.cpc.2007.02.075} {\bibfield
  {journal} {\bibinfo  {journal} {Computer Physics Communications}\ }\textbf
  {\bibinfo {volume} {177}},\ \bibinfo {pages} {14 } (\bibinfo {year}
  {2007})},\ \bibinfo {note} {proceedings of the Conference on Computational
  Physics 2006}\BibitemShut {NoStop}%
\bibitem [{\citenamefont {Blum}\ \emph {et~al.}(2009)\citenamefont {Blum},
  \citenamefont {Gehrke}, \citenamefont {Hanke}, \citenamefont {Havu},
  \citenamefont {Havu}, \citenamefont {Ren}, \citenamefont {Reuter},\ and\
  \citenamefont {Scheffler}}]{BLUM20092175}%
  \BibitemOpen
  \bibfield  {author} {\bibinfo {author} {\bibfnamefont {V.}~\bibnamefont
  {Blum}}, \bibinfo {author} {\bibfnamefont {R.}~\bibnamefont {Gehrke}},
  \bibinfo {author} {\bibfnamefont {F.}~\bibnamefont {Hanke}}, \bibinfo
  {author} {\bibfnamefont {P.}~\bibnamefont {Havu}}, \bibinfo {author}
  {\bibfnamefont {V.}~\bibnamefont {Havu}}, \bibinfo {author} {\bibfnamefont
  {X.}~\bibnamefont {Ren}}, \bibinfo {author} {\bibfnamefont {K.}~\bibnamefont
  {Reuter}},\ and\ \bibinfo {author} {\bibfnamefont {M.}~\bibnamefont
  {Scheffler}},\ }\href
  {https://doi.org/https://doi.org/10.1016/j.cpc.2009.06.022} {\bibfield
  {journal} {\bibinfo  {journal} {Computer Physics Communications}\ }\textbf
  {\bibinfo {volume} {180}},\ \bibinfo {pages} {2175 } (\bibinfo {year}
  {2009})}\BibitemShut {NoStop}%
\bibitem [{\citenamefont {Hutter}\ \emph {et~al.}(2014)\citenamefont {Hutter},
  \citenamefont {Iannuzzi}, \citenamefont {Schiffmann},\ and\ \citenamefont
  {VandeVondele}}]{hutter2014cp2k}%
  \BibitemOpen
  \bibfield  {author} {\bibinfo {author} {\bibfnamefont {J.}~\bibnamefont
  {Hutter}}, \bibinfo {author} {\bibfnamefont {M.}~\bibnamefont {Iannuzzi}},
  \bibinfo {author} {\bibfnamefont {F.}~\bibnamefont {Schiffmann}},\ and\
  \bibinfo {author} {\bibfnamefont {J.}~\bibnamefont {VandeVondele}},\
  }\href@noop {} {\bibfield  {journal} {\bibinfo  {journal} {Wiley
  Interdisciplinary Reviews: Computational Molecular Science}\ }\textbf
  {\bibinfo {volume} {4}},\ \bibinfo {pages} {15} (\bibinfo {year}
  {2014})}\BibitemShut {NoStop}%
\bibitem [{qua()}]{quantumwise}%
  \BibitemOpen
  \href@noop {} {}\bibinfo {note} {The information of Atomistix ToolKit can be
  found on a website (http://www.quantumwise.com/).}\BibitemShut {Stop}%
\bibitem [{\citenamefont {Ozaki}(2003)}]{Ozaki}%
  \BibitemOpen
  \bibfield  {author} {\bibinfo {author} {\bibfnamefont {T.}~\bibnamefont
  {Ozaki}},\ }\href {https://doi.org/10.1103/PhysRevB.67.155108} {\bibfield
  {journal} {\bibinfo  {journal} {Phys. Rev. B}\ }\textbf {\bibinfo {volume}
  {67}},\ \bibinfo {pages} {155108} (\bibinfo {year} {2003})}\BibitemShut
  {NoStop}%
\bibitem [{\citenamefont {Ga\'al-Nagy}(2008)}]{PhysRevB.77.024309}%
  \BibitemOpen
  \bibfield  {author} {\bibinfo {author} {\bibfnamefont {K.}~\bibnamefont
  {Ga\'al-Nagy}},\ }\bibfield  {title} {\bibinfo {title} {Accelerating ab
  initio calculation of phonon dispersion curves: $\mathbf{q}$-point
  convergence},\ }\href {https://doi.org/10.1103/PhysRevB.77.024309} {\bibfield
   {journal} {\bibinfo  {journal} {Phys. Rev. B}\ }\textbf {\bibinfo {volume}
  {77}},\ \bibinfo {pages} {024309} (\bibinfo {year} {2008})}\BibitemShut
  {NoStop}%
\bibitem [{\citenamefont {Zhou}\ \emph {et~al.}(2019)\citenamefont {Zhou},
  \citenamefont {Sadigh}, \citenamefont {\AA{}berg}, \citenamefont {Xia},\ and\
  \citenamefont {Ozoli\ifmmode \mbox{\c{n}}\else
  \c{n}\fi{}\ifmmode~\check{s}\else \v{s}\fi{}}}]{PhysRevB.100.184309}%
  \BibitemOpen
  \bibfield  {author} {\bibinfo {author} {\bibfnamefont {F.}~\bibnamefont
  {Zhou}}, \bibinfo {author} {\bibfnamefont {B.}~\bibnamefont {Sadigh}},
  \bibinfo {author} {\bibfnamefont {D.}~\bibnamefont {\AA{}berg}}, \bibinfo
  {author} {\bibfnamefont {Y.}~\bibnamefont {Xia}},\ and\ \bibinfo {author}
  {\bibfnamefont {V.}~\bibnamefont {Ozoli\ifmmode \mbox{\c{n}}\else
  \c{n}\fi{}\ifmmode~\check{s}\else \v{s}\fi{}}},\ }\href
  {https://doi.org/10.1103/PhysRevB.100.184309} {\bibfield  {journal} {\bibinfo
   {journal} {Phys. Rev. B}\ }\textbf {\bibinfo {volume} {100}},\ \bibinfo
  {pages} {184309} (\bibinfo {year} {2019})}\BibitemShut {NoStop}%
\bibitem [{\citenamefont {Wang}\ \emph {et~al.}(2010)\citenamefont {Wang},
  \citenamefont {Wang}, \citenamefont {Wang}, \citenamefont {Mei},
  \citenamefont {Shang}, \citenamefont {Chen},\ and\ \citenamefont
  {Liu}}]{mixed}%
  \BibitemOpen
  \bibfield  {author} {\bibinfo {author} {\bibfnamefont {Y.}~\bibnamefont
  {Wang}}, \bibinfo {author} {\bibfnamefont {J.~J.}\ \bibnamefont {Wang}},
  \bibinfo {author} {\bibfnamefont {W.~Y.}\ \bibnamefont {Wang}}, \bibinfo
  {author} {\bibfnamefont {Z.~G.}\ \bibnamefont {Mei}}, \bibinfo {author}
  {\bibfnamefont {S.~L.}\ \bibnamefont {Shang}}, \bibinfo {author}
  {\bibfnamefont {L.~Q.}\ \bibnamefont {Chen}},\ and\ \bibinfo {author}
  {\bibfnamefont {Z.~K.}\ \bibnamefont {Liu}},\ }\href@noop {} {\bibfield
  {journal} {\bibinfo  {journal} {J. Phys.: Condens. Matter}\ }\textbf
  {\bibinfo {volume} {22}},\ \bibinfo {pages} {202201} (\bibinfo {year}
  {2010})}\BibitemShut {NoStop}%
\bibitem [{\citenamefont {Wang}\ \emph {et~al.}(2012)\citenamefont {Wang},
  \citenamefont {Shang}, \citenamefont {Liu},\ and\ \citenamefont
  {Chen}}]{PhysRevB.85.224303}%
  \BibitemOpen
  \bibfield  {author} {\bibinfo {author} {\bibfnamefont {Y.}~\bibnamefont
  {Wang}}, \bibinfo {author} {\bibfnamefont {S.}~\bibnamefont {Shang}},
  \bibinfo {author} {\bibfnamefont {Z.-K.}\ \bibnamefont {Liu}},\ and\ \bibinfo
  {author} {\bibfnamefont {L.-Q.}\ \bibnamefont {Chen}},\ }\href
  {https://doi.org/10.1103/PhysRevB.85.224303} {\bibfield  {journal} {\bibinfo
  {journal} {Phys. Rev. B}\ }\textbf {\bibinfo {volume} {85}},\ \bibinfo
  {pages} {224303} (\bibinfo {year} {2012})}\BibitemShut {NoStop}%
\bibitem [{Ste()}]{Stengel}%
  \BibitemOpen
  \href@noop {} {}\bibinfo {note} {M. Royo, K. R. Hahn, and M. Stengel,
  arXiv:2004.08875 (2020).}\BibitemShut {Stop}%
\bibitem [{\citenamefont {Cochran}\ and\ \citenamefont
  {Cowley}(1962)}]{Cochran}%
  \BibitemOpen
  \bibfield  {author} {\bibinfo {author} {\bibfnamefont {W.}~\bibnamefont
  {Cochran}}\ and\ \bibinfo {author} {\bibfnamefont {R.~A.}\ \bibnamefont
  {Cowley}},\ }\href {https://doi.org/10.1016/0022-3697(62)90084-7} {\bibfield
  {journal} {\bibinfo  {journal} {J. Phys. Chem. Solids}\ }\textbf {\bibinfo
  {volume} {23}},\ \bibinfo {pages} {447} (\bibinfo {year} {1962})}\BibitemShut
  {NoStop}%
\bibitem [{\citenamefont {Parlinski}\ \emph {et~al.}(1998)\citenamefont
  {Parlinski}, \citenamefont {Li},\ and\ \citenamefont
  {Kawazoe}}]{PhysRevLett.81.3298}%
  \BibitemOpen
  \bibfield  {author} {\bibinfo {author} {\bibfnamefont {K.}~\bibnamefont
  {Parlinski}}, \bibinfo {author} {\bibfnamefont {Z.~Q.}\ \bibnamefont {Li}},\
  and\ \bibinfo {author} {\bibfnamefont {Y.}~\bibnamefont {Kawazoe}},\ }\href
  {https://doi.org/10.1103/PhysRevLett.81.3298} {\bibfield  {journal} {\bibinfo
   {journal} {Phys. Rev. Lett.}\ }\textbf {\bibinfo {volume} {81}},\ \bibinfo
  {pages} {3298} (\bibinfo {year} {1998})}\BibitemShut {NoStop}%
\bibitem [{\citenamefont {Wang}\ \emph {et~al.}(2014)\citenamefont {Wang},
  \citenamefont {Chen},\ and\ \citenamefont {Liu}}]{wang2014yphon}%
  \BibitemOpen
  \bibfield  {author} {\bibinfo {author} {\bibfnamefont {Y.}~\bibnamefont
  {Wang}}, \bibinfo {author} {\bibfnamefont {L.-Q.}\ \bibnamefont {Chen}},\
  and\ \bibinfo {author} {\bibfnamefont {Z.-K.}\ \bibnamefont {Liu}},\
  }\href@noop {} {\bibfield  {journal} {\bibinfo  {journal} {Computer Physics
  Communications}\ }\textbf {\bibinfo {volume} {185}},\ \bibinfo {pages} {2950}
  (\bibinfo {year} {2014})}\BibitemShut {NoStop}%
\bibitem [{\citenamefont {Li}\ \emph {et~al.}(2014)\citenamefont {Li},
  \citenamefont {Carrete}, \citenamefont {Katcho},\ and\ \citenamefont
  {Mingo}}]{li2014shengbte}%
  \BibitemOpen
  \bibfield  {author} {\bibinfo {author} {\bibfnamefont {W.}~\bibnamefont
  {Li}}, \bibinfo {author} {\bibfnamefont {J.}~\bibnamefont {Carrete}},
  \bibinfo {author} {\bibfnamefont {N.~A.}\ \bibnamefont {Katcho}},\ and\
  \bibinfo {author} {\bibfnamefont {N.}~\bibnamefont {Mingo}},\ }\href@noop {}
  {\bibfield  {journal} {\bibinfo  {journal} {Computer Physics Communications}\
  }\textbf {\bibinfo {volume} {185}},\ \bibinfo {pages} {1747} (\bibinfo {year}
  {2014})}\BibitemShut {NoStop}%
\bibitem [{\citenamefont {Chernatynskiy}\ and\ \citenamefont
  {Phillpot}(2015)}]{chernatynskiy2015phonon}%
  \BibitemOpen
  \bibfield  {author} {\bibinfo {author} {\bibfnamefont {A.}~\bibnamefont
  {Chernatynskiy}}\ and\ \bibinfo {author} {\bibfnamefont {S.~R.}\ \bibnamefont
  {Phillpot}},\ }\href@noop {} {\bibfield  {journal} {\bibinfo  {journal}
  {Computer Physics Communications}\ }\textbf {\bibinfo {volume} {192}},\
  \bibinfo {pages} {196} (\bibinfo {year} {2015})}\BibitemShut {NoStop}%
\bibitem [{\citenamefont {Togo}\ \emph {et~al.}(2008)\citenamefont {Togo},
  \citenamefont {Oba},\ and\ \citenamefont {Tanaka}}]{togo2008first}%
  \BibitemOpen
  \bibfield  {author} {\bibinfo {author} {\bibfnamefont {A.}~\bibnamefont
  {Togo}}, \bibinfo {author} {\bibfnamefont {F.}~\bibnamefont {Oba}},\ and\
  \bibinfo {author} {\bibfnamefont {I.}~\bibnamefont {Tanaka}},\ }\href@noop {}
  {\bibfield  {journal} {\bibinfo  {journal} {Physical Review B}\ }\textbf
  {\bibinfo {volume} {78}},\ \bibinfo {pages} {134106} (\bibinfo {year}
  {2008})}\BibitemShut {NoStop}%
\bibitem [{\citenamefont {Tadano}\ \emph {et~al.}(2014)\citenamefont {Tadano},
  \citenamefont {Gohda},\ and\ \citenamefont
  {Tsuneyuki}}]{tadano2014anharmonic}%
  \BibitemOpen
  \bibfield  {author} {\bibinfo {author} {\bibfnamefont {T.}~\bibnamefont
  {Tadano}}, \bibinfo {author} {\bibfnamefont {Y.}~\bibnamefont {Gohda}},\ and\
  \bibinfo {author} {\bibfnamefont {S.}~\bibnamefont {Tsuneyuki}},\ }\href@noop
  {} {\bibfield  {journal} {\bibinfo  {journal} {Journal of Physics: Condensed
  Matter}\ }\textbf {\bibinfo {volume} {26}},\ \bibinfo {pages} {225402}
  (\bibinfo {year} {2014})}\BibitemShut {NoStop}%
\bibitem [{ope()}]{openmx}%
  \BibitemOpen
  \href@noop {} {}\bibinfo {note} {The code, OpenMX, pseudo-atomic basis
  functions, and pseudopotentials are available on a website
  (http://www.openmx-square.org/).}\BibitemShut {Stop}%
\bibitem [{\citenamefont {Perdew}\ \emph {et~al.}(1996)\citenamefont {Perdew},
  \citenamefont {Burke},\ and\ \citenamefont {Ernzerhof}}]{GGA}%
  \BibitemOpen
  \bibfield  {author} {\bibinfo {author} {\bibfnamefont {J.~P.}\ \bibnamefont
  {Perdew}}, \bibinfo {author} {\bibfnamefont {K.}~\bibnamefont {Burke}},\ and\
  \bibinfo {author} {\bibfnamefont {M.}~\bibnamefont {Ernzerhof}},\ }\href
  {https://doi.org/10.1103/PhysRevLett.77.3865} {\bibfield  {journal} {\bibinfo
   {journal} {Phys. Rev. Lett.}\ }\textbf {\bibinfo {volume} {77}},\ \bibinfo
  {pages} {3865} (\bibinfo {year} {1996})}\BibitemShut {NoStop}%
\bibitem [{\citenamefont {Theurich}\ and\ \citenamefont
  {Hill}(2001)}]{Theurich}%
  \BibitemOpen
  \bibfield  {author} {\bibinfo {author} {\bibfnamefont {G.}~\bibnamefont
  {Theurich}}\ and\ \bibinfo {author} {\bibfnamefont {N.~A.}\ \bibnamefont
  {Hill}},\ }\href {https://doi.org/10.1103/PhysRevB.64.073106} {\bibfield
  {journal} {\bibinfo  {journal} {Phys. Rev. B}\ }\textbf {\bibinfo {volume}
  {64}},\ \bibinfo {pages} {073106} (\bibinfo {year} {2001})}\BibitemShut
  {NoStop}%
\bibitem [{\citenamefont {Morrison}\ \emph {et~al.}(1993)\citenamefont
  {Morrison}, \citenamefont {Bylander},\ and\ \citenamefont
  {Kleinman}}]{Morrison}%
  \BibitemOpen
  \bibfield  {author} {\bibinfo {author} {\bibfnamefont {I.}~\bibnamefont
  {Morrison}}, \bibinfo {author} {\bibfnamefont {D.~M.}\ \bibnamefont
  {Bylander}},\ and\ \bibinfo {author} {\bibfnamefont {L.}~\bibnamefont
  {Kleinman}},\ }\href {https://doi.org/10.1103/PhysRevB.47.6728} {\bibfield
  {journal} {\bibinfo  {journal} {Phys. Rev. B}\ }\textbf {\bibinfo {volume}
  {47}},\ \bibinfo {pages} {6728} (\bibinfo {year} {1993})}\BibitemShut
  {NoStop}%
\bibitem [{\citenamefont {Raunio}\ \emph {et~al.}(1969)\citenamefont {Raunio},
  \citenamefont {Almqvist},\ and\ \citenamefont {Stedman}}]{PhysRev.178.1496}%
  \BibitemOpen
  \bibfield  {author} {\bibinfo {author} {\bibfnamefont {G.}~\bibnamefont
  {Raunio}}, \bibinfo {author} {\bibfnamefont {L.}~\bibnamefont {Almqvist}},\
  and\ \bibinfo {author} {\bibfnamefont {R.}~\bibnamefont {Stedman}},\ }\href
  {https://doi.org/10.1103/PhysRev.178.1496} {\bibfield  {journal} {\bibinfo
  {journal} {Phys. Rev.}\ }\textbf {\bibinfo {volume} {178}},\ \bibinfo {pages}
  {1496} (\bibinfo {year} {1969})}\BibitemShut {NoStop}%
\bibitem [{\citenamefont {Marcondes}\ \emph {et~al.}(2018)\citenamefont
  {Marcondes}, \citenamefont {Wentzcovitch},\ and\ \citenamefont
  {Assali}}]{MARCONDES201811}%
  \BibitemOpen
  \bibfield  {author} {\bibinfo {author} {\bibfnamefont {M.~L.}\ \bibnamefont
  {Marcondes}}, \bibinfo {author} {\bibfnamefont {R.~M.}\ \bibnamefont
  {Wentzcovitch}},\ and\ \bibinfo {author} {\bibfnamefont {L.~V.}\ \bibnamefont
  {Assali}},\ }\href
  {https://doi.org/https://doi.org/10.1016/j.ssc.2018.01.008} {\bibfield
  {journal} {\bibinfo  {journal} {Solid State Communications}\ }\textbf
  {\bibinfo {volume} {273}},\ \bibinfo {pages} {11 } (\bibinfo {year}
  {2018})}\BibitemShut {NoStop}%
\bibitem [{\citenamefont {Kern}\ \emph {et~al.}(1999)\citenamefont {Kern},
  \citenamefont {Kresse},\ and\ \citenamefont {Hafner}}]{PhysRevB.59.8551}%
  \BibitemOpen
  \bibfield  {author} {\bibinfo {author} {\bibfnamefont {G.}~\bibnamefont
  {Kern}}, \bibinfo {author} {\bibfnamefont {G.}~\bibnamefont {Kresse}},\ and\
  \bibinfo {author} {\bibfnamefont {J.}~\bibnamefont {Hafner}},\ }\href
  {https://doi.org/10.1103/PhysRevB.59.8551} {\bibfield  {journal} {\bibinfo
  {journal} {Phys. Rev. B}\ }\textbf {\bibinfo {volume} {59}},\ \bibinfo
  {pages} {8551} (\bibinfo {year} {1999})}\BibitemShut {NoStop}%
\bibitem [{\citenamefont {Umari}\ and\ \citenamefont
  {Pasquarello}(2002)}]{PhysRevLett.89.157602}%
  \BibitemOpen
  \bibfield  {author} {\bibinfo {author} {\bibfnamefont {P.}~\bibnamefont
  {Umari}}\ and\ \bibinfo {author} {\bibfnamefont {A.}~\bibnamefont
  {Pasquarello}},\ }\href {https://doi.org/10.1103/PhysRevLett.89.157602}
  {\bibfield  {journal} {\bibinfo  {journal} {Phys. Rev. Lett.}\ }\textbf
  {\bibinfo {volume} {89}},\ \bibinfo {pages} {157602} (\bibinfo {year}
  {2002})}\BibitemShut {NoStop}%
\bibitem [{\citenamefont {King-Smith}\ and\ \citenamefont
  {Vanderbilt}(1993)}]{PhysRevB.47.1651}%
  \BibitemOpen
  \bibfield  {author} {\bibinfo {author} {\bibfnamefont {R.~D.}\ \bibnamefont
  {King-Smith}}\ and\ \bibinfo {author} {\bibfnamefont {D.}~\bibnamefont
  {Vanderbilt}},\ }\href {https://doi.org/10.1103/PhysRevB.47.1651} {\bibfield
  {journal} {\bibinfo  {journal} {Phys. Rev. B}\ }\textbf {\bibinfo {volume}
  {47}},\ \bibinfo {pages} {1651} (\bibinfo {year} {1993})}\BibitemShut
  {NoStop}%
\bibitem [{\citenamefont {Gonze}\ \emph {et~al.}(2016)\citenamefont {Gonze},
  \citenamefont {Jollet}, \citenamefont {Araujo}, \citenamefont {Adams},
  \citenamefont {Amadon}, \citenamefont {Applencourt}, \citenamefont {Audouze}
  \emph {et~al.}}]{GONZE2016106}%
  \BibitemOpen
  \bibfield  {author} {\bibinfo {author} {\bibfnamefont {X.}~\bibnamefont
  {Gonze}}, \bibinfo {author} {\bibfnamefont {F.}~\bibnamefont {Jollet}},
  \bibinfo {author} {\bibfnamefont {F.~A.}\ \bibnamefont {Araujo}}, \bibinfo
  {author} {\bibfnamefont {D.}~\bibnamefont {Adams}}, \bibinfo {author}
  {\bibfnamefont {B.}~\bibnamefont {Amadon}}, \bibinfo {author} {\bibfnamefont
  {T.}~\bibnamefont {Applencourt}}, \bibinfo {author} {\bibfnamefont
  {C.}~\bibnamefont {Audouze}}, \emph {et~al.},\ }\href
  {https://doi.org/https://doi.org/10.1016/j.cpc.2016.04.003} {\bibfield
  {journal} {\bibinfo  {journal} {Computer Physics Communications}\ }\textbf
  {\bibinfo {volume} {205}},\ \bibinfo {pages} {106 } (\bibinfo {year}
  {2016})}\BibitemShut {NoStop}%
\bibitem [{\citenamefont {Hamann}(2013)}]{PhysRevB.88.085117}%
  \BibitemOpen
  \bibfield  {author} {\bibinfo {author} {\bibfnamefont {D.~R.}\ \bibnamefont
  {Hamann}},\ }\href {https://doi.org/10.1103/PhysRevB.88.085117} {\bibfield
  {journal} {\bibinfo  {journal} {Phys. Rev. B}\ }\textbf {\bibinfo {volume}
  {88}},\ \bibinfo {pages} {085117} (\bibinfo {year} {2013})}\BibitemShut
  {NoStop}%
\bibitem [{\citenamefont {Lucas~Webster}\ and\ \citenamefont
  {Yan}(2018)}]{Lucas}%
  \BibitemOpen
  \bibfield  {author} {\bibinfo {author} {\bibfnamefont {L.~L.}\ \bibnamefont
  {Lucas~Webster}}\ and\ \bibinfo {author} {\bibfnamefont {J.-A.}\ \bibnamefont
  {Yan}},\ }\href@noop {} {\bibfield  {journal} {\bibinfo  {journal} {Phys.
  Chem. Chem. Phys.}\ }\textbf {\bibinfo {volume} {20}},\ \bibinfo {pages}
  {23546} (\bibinfo {year} {2018})}\BibitemShut {NoStop}%
\bibitem [{\citenamefont {Cocemasov}\ \emph {et~al.}(2013)\citenamefont
  {Cocemasov}, \citenamefont {Nika},\ and\ \citenamefont
  {Balandin}}]{PhysRevB.88.035428}%
  \BibitemOpen
  \bibfield  {author} {\bibinfo {author} {\bibfnamefont {A.~I.}\ \bibnamefont
  {Cocemasov}}, \bibinfo {author} {\bibfnamefont {D.~L.}\ \bibnamefont
  {Nika}},\ and\ \bibinfo {author} {\bibfnamefont {A.~A.}\ \bibnamefont
  {Balandin}},\ }\href {https://doi.org/10.1103/PhysRevB.88.035428} {\bibfield
  {journal} {\bibinfo  {journal} {Phys. Rev. B}\ }\textbf {\bibinfo {volume}
  {88}},\ \bibinfo {pages} {035428} (\bibinfo {year} {2013})}\BibitemShut
  {NoStop}%
\bibitem [{\citenamefont {Song}\ \emph {et~al.}(2019)\citenamefont {Song},
  \citenamefont {Liu},\ and\ \citenamefont {Zhang}}]{SONG20192628}%
  \BibitemOpen
  \bibfield  {author} {\bibinfo {author} {\bibfnamefont {H.-Q.}\ \bibnamefont
  {Song}}, \bibinfo {author} {\bibfnamefont {Z.}~\bibnamefont {Liu}},\ and\
  \bibinfo {author} {\bibfnamefont {D.-B.}\ \bibnamefont {Zhang}},\ }\href
  {https://doi.org/https://doi.org/10.1016/j.physleta.2019.05.025} {\bibfield
  {journal} {\bibinfo  {journal} {Physics Letters A}\ }\textbf {\bibinfo
  {volume} {383}},\ \bibinfo {pages} {2628 } (\bibinfo {year}
  {2019})}\BibitemShut {NoStop}%
\bibitem [{\citenamefont {Ramnani}\ \emph {et~al.}(2017)\citenamefont
  {Ramnani}, \citenamefont {Neupane}, \citenamefont {Ge}, \citenamefont
  {Balandin}, \citenamefont {Lake},\ and\ \citenamefont
  {Mulchandani}}]{RAMNANI2017302}%
  \BibitemOpen
  \bibfield  {author} {\bibinfo {author} {\bibfnamefont {P.}~\bibnamefont
  {Ramnani}}, \bibinfo {author} {\bibfnamefont {M.~R.}\ \bibnamefont
  {Neupane}}, \bibinfo {author} {\bibfnamefont {S.}~\bibnamefont {Ge}},
  \bibinfo {author} {\bibfnamefont {A.~A.}\ \bibnamefont {Balandin}}, \bibinfo
  {author} {\bibfnamefont {R.~K.}\ \bibnamefont {Lake}},\ and\ \bibinfo
  {author} {\bibfnamefont {A.}~\bibnamefont {Mulchandani}},\ }\href
  {https://doi.org/https://doi.org/10.1016/j.carbon.2017.07.064} {\bibfield
  {journal} {\bibinfo  {journal} {Carbon}\ }\textbf {\bibinfo {volume} {123}},\
  \bibinfo {pages} {302 } (\bibinfo {year} {2017})}\BibitemShut {NoStop}%
\bibitem [{\citenamefont {Grimme}(2006)}]{grimme2006semiempirical}%
  \BibitemOpen
  \bibfield  {author} {\bibinfo {author} {\bibfnamefont {S.}~\bibnamefont
  {Grimme}},\ }\href@noop {} {\bibfield  {journal} {\bibinfo  {journal}
  {Journal of computational chemistry}\ }\textbf {\bibinfo {volume} {27}},\
  \bibinfo {pages} {1787} (\bibinfo {year} {2006})}\BibitemShut {NoStop}%
\end{thebibliography}%

\end{document}